\documentclass[
aps
,groupedaddress
,amsfonts,amsmath,floatfix,eqsecnum
,nofootinbib
,twocolumn
]{revtex4}

\usepackage{mathtools}
\usepackage{hyperref}

\usepackage[utf8]{inputenc}
\usepackage{epsfig,graphicx}
\usepackage{bm}
\usepackage{amsmath,amsfonts}
\usepackage{enumerate}
\usepackage{ytableau}
\usepackage{cancel}
\usepackage{multirow}
\usepackage{mathptmx}      
\usepackage{latexsym}
\usepackage{amstext}
\usepackage{array}
\usepackage[T1]{fontenc} 

\DeclareMathSymbol{\gtrless} {\mathrel}{AMSa}{"3F}

\newcommand{\ignore}[1]{}
\newcommand{\nec}[1]{\eqref{eq:#1}}

\newcommand{\eq}[1]{eq. \eqref{eq:#1}}
\newcommand{\eqs}[1]{eqs. \eqref{eq:#1}}
\newcommand{\Eq}[1]{Eq. \eqref{eq:#1}}

\newcommand{\be}{\begin{equation}}
\newcommand{\ee}{\end{equation}}
\def\bes#1\ees{%
  \begin{equation}
    \begin{split}
      #1
    \end{split}
  \end{equation}
}
\def\bs#1\es{%
    \begin{split}
      #1
    \end{split}
}
\newcommand{\C}{\mathbb{C}}

\newcommand{\Z}{\mathbb{Z}}

\def\slashchar#1{\setbox0=\hbox{$#1$}
   \dimen0=\wd0 \setbox1=\hbox{/} \dimen1=\wd1
   \ifdim\dimen0>\dimen1 \rlap{\hbox to \dimen0{\hfil/\hfil}} #1
   \else  \rlap{\hbox to \dimen1{\hfil$#1$\hfil}} / \fi}

\newcommand{\fin}{\hbox{~$\diamondsuit$}}

\newcommand{\ben}{\begin{enumerate}}
\newcommand{\een}{\end{enumerate}}

\newcommand{\ds}{\displaystyle}

\DeclareMathOperator{\Tr}{Tr}

\newcommand{\cA}{{\mathcal A }}
\newcommand{\cB}{{\mathcal B }}

\newcommand{\cF}{{\mathcal F }}
\newcommand{\cG}{{\mathcal G }}
\newcommand{\cH}{{\mathcal H }}

\newcommand{\cL}{{\mathcal L }}

\newcommand{\cS}{{\mathcal S }}

\newcommand{\cV}{{\mathcal V }}
\newcommand{\cW}{{\mathcal W }}

\newcommand{\esp}[1]{\langle #1 \rangle}

\newcommand{\miem}[1]{\textbf{\textsf{#1}}}

\newcommand{\PM}[1]{ \begin{pmatrix} #1 \end{pmatrix} }
\newcommand{\Oplus}{\ensuremath{\vcenter{\hbox{\scalebox{1.1}{$\bigoplus$}}}}}

\newcommand{\OcU}{\Omega\cdot U}
\newcommand{\malpha}{t}
\newcommand{\mbeta}{r}

\newcommand{\ket}[1]{ | #1 \rangle }

\renewcommand{\Omega}{\varOmega}
\newcommand{\tT}{\tilde{T}}
\newcommand{\tU}{\tilde{U}}

\newcommand{\qed}{ \hfill $\square$ }
\renewcommand{\fin}{ \hfill $\diamondsuit$ }
\newcommand{\proof}{ \miem{Proof} }
\newcounter{ctth} \renewcommand{\newtheorem}[1]{
  \refstepcounter{ctth}\label{#1} \miem{Theorem \thectth } }
\newcounter{ctlm} \newcommand{\newlemma}[1]{ \refstepcounter{ctlm}\label{#1}
  \miem{Lemma \thectlm } }
\newcounter{ctpr} \newcommand{\newproposition}[1]{
  \refstepcounter{ctpr}\label{#1} \miem{Proposition \thectpr } }
\newcounter{ctex} \newcommand{\newexample}[1]{
  \refstepcounter{ctex}\label{#1} \miem{Example \thectex } }
\newcounter{ctpc} \newcommand{\newparticularcase}[1]{
  \refstepcounter{ctpc}\label{#1} \miem{Particular case \thectpc } }
\newcounter{ctdf} \newcommand{\newdefinition}[1]{
  \refstepcounter{ctdf}\label{#1} \miem{Definition \thectdf } }
\newcounter{ctcor} \newcommand{\newcorollary}[1]{
  \refstepcounter{ctcor}\label{#1} \miem{Corollary \thectcor } }

\newcommand{\mcirc}{\ensuremath{\vcenter{\hbox{\scalebox{0.7}{$\circ$}}}}}
\newcommand{\supop}[1]{\ensuremath{\overbracket[0.1ex][0.3ex]{#1}}}

\begin{document}

\title{\textsf{
    Reduction of unitary operators, quantum graphs and quantum channels
  }}

\author{L. L. Salcedo}
\email{salcedo@ugr.es}

\affiliation{Departamento de F\'{\i}sica At\'omica, Molecular y Nuclear and \\
  Instituto Carlos I de F\'{\i}sica Te\'orica y Computacional, \\ Universidad
  de Granada, E-18071 Granada, Spain.
}

\date{\today}

\begin{abstract}
  Given a unitary operator in a finite dimensional complex Hilbert space, its
  unitary reduction to a subspace is defined.  The application to quantum
  graphs is discussed.  It is shown how the reduction allows to generate the
  scattering matrices of new quantum graphs from assembling of simpler graphs.
  The reduction of quantum channels is also defined.  The implementation of
  the quantum gates corresponding to the reduced unitary operator is
  investigated, although no explicit construction is presented. The situation
  is different for the reduction of quantum channels for which explicit
  implementations are given.
\end{abstract}



\maketitle
\flushbottom
\setlength{\unitlength}{1mm}

\tableofcontents

\newpage

\section{ Introduction }
\label{sec:0}

In this work we introduce the concept of reduction of a unitary operator to
produce another unitary operator which acts in a subspace. The concept
originates in the context of quantum graphs. Quantum graphs describe a quantum
particle, which may have spin, or be relativistic, moving on a net of lines
connected at nodes, forming a graph. The graph may be compact, the Hamiltonian
having only a discrete spectrum, or non compact, so that the particle may go
towards (or come from) infinity along several leads. In the latter case a
scattering matrix may be defined under suitable assumptions ensuring that the
particle moves freely in the asymptotic region. Since the kinematics is
trivial in the asymptotic region, when two graphs are joined to form a larger
graph, the scattering matrix of the latter is determined by the scattering
matrices of the components (plus information on the junctions, namely, their
lengths). The resulting formula is an instance of reduction of unitary
operators. \ignore{The adopted name of ``reduction'' comes from this context and
also because it loosely recalls (a nonlinear version of) the reduction of a
tensor to produce a tensor of lower rank.}

The reduction of finite-dimensional unitary operators is defined and
analyzed here from the mathematical point of view. It is shown that it exists
with no restrictions on the unitary operators or the subspaces. The
reduction may be introduced through a set of equations or as the sum of a
series. Both approaches are discussed in Sec. \ref{sec:1} and several
properties of the reduction are uncovered, providing the corresponding
mathematical proofs.  Sec. \ref{sec:3} introduces quantum graphs. The relation
of the reduction of graphs, or rather their scattering matrices, with the
reduction of unitary operators is established.  In the first part of
Sec. \ref{sec:2} we investigate the reduction of general, i.e., non-unitary,
operators within the two approaches (equations and series). Not surprisingly,
at variance with the unitary case, the operators should fulfill some specific
requirements for the existence of the reduction. In the second part of
Sec. \ref{sec:2} the concept of reduction is extended to quantum
channels. Necessary and sufficient conditions on the quantum channels are
established for the reduction to exist.  Finally, in Sec. \ref{sec:4} we
consider the possible implementation of the unitary reduction in the context
of quantum computation. Certainly every finite unitary quantum gate can be
constructed using circuits but in general this requires a detailed knowledge
of the corresponding unitary operator. What is studied here is whether a
circuit implementing the reduction can be produced given the unitary
operator to be reduced as a black box. The problem is related to the
introduction of loops in a quantum circuit. Our analysis does not find a
positive answer, but it does for the related problem of the reduction of
quantum channels.

\section{  Reduction of a unitary operator to a subspace }
\label{sec:1}

\subsection{  Definitions and main theorem }

In this work all abstract vector spaces considered are complex and finite
dimensional, unless otherwise specified. In the case of Hilbert spaces
$\oplus$ denotes the direct sum of orthogonal subspaces, $\subseteq$ denotes
subspace, and $\ominus$ denotes orthogonal complement.

\newlemma{lm:3} ~Let $\cW \subseteq \cA \oplus \cB $ be a relation between
Hilbert spaces. Then
\be
\cW = W \cA \oplus \cW \cap \cB
,
\ee
where $W$ is the orthogonal projector operator onto the subspace $\cW$.

\proof
~Let $\varphi\in\cW$. Then
$\varphi \in \cB \iff \varphi \perp \cA \iff \varphi \perp W\cA$. This implies
$\cW\cap \cB = \cW \ominus W \cA$. \qed

\newtheorem{th:1} ~Let $U:\cH\to\cF$ and $\Omega:\cF_1\to\cH_1$ be unitary
operators between Hilbert spaces with $\cF_1\subseteq \cF$ and
$\cH_1\subseteq \cH$.  Let $\cH_0 := \cH \ominus \cH_1$ and
$\cF_0 := \cF \ominus \cF_1$. Then the equations
\be
\phi_0 + \phi_1 = U ( \psi_0 + \psi_1 ),
\qquad
\psi_1 = \Omega \phi_1
,
\label{eq:1.1}
\ee
with unknowns $\psi_1\in\cH_1$, $\phi_0\in\cF_0$ and $\phi_1\in\cF_1$, have a
solution for any given $\psi_0\in\cH_0$. Furthermore the solution is unique
for $\phi_0$ and $\|\phi_0\| = \|\psi_0\|$.

\proof ~Let $H_j$ and $F_j$, for $j\in\{0,1\}$, denote the orthogonal
projector operators onto $\cH_j$ and $\cF_j$, respectively. Then, upon
elimination of $\psi_1$ in favor of $\phi_1$, the conditions can be expressed
as
\bes
\phi_0 &= F_0 U \psi_0 + F_0 U \Omega \phi_1
,
\\
\phi_1 &= F_1 U \psi_0 + F_1 U \Omega\phi_1
,
\label{eq:1.3}
\ees
and the second equation is equivalent to
\be
F_1 U \psi_0  = (I - F_1 U \Omega)\phi_1 
\label{eq:1.3b}
.
\ee
This equation can be solved for $\phi_1$, for arbitrary $\psi_0\in\cH_0$, if
and only if $F_1 U \cH_0 \subseteq (I - F_1 U \Omega)\cF_1$.

Let
\be
\cV_1 :=\{ \varphi_1 \in \cF_1 ~|~ U \Omega \varphi_1 = \varphi_1
\}
.
\label{eq:2.5}
\ee
Let us show that $\cV_1$ is the kernel of the operator
$(I - F_1 U \Omega):\cF_1 \to \cF_1$.  Clearly,
$\varphi_1 \in \cV_1 \implies (I - F_1 U \Omega) \varphi_1 = 0$. And vice
versa, $F_1 U \Omega \varphi_1 = \varphi_1$ implies that
$U\Omega\varphi_1 = \varphi_1 + \varphi_0$ with $\varphi_0\in\cF_0$.  However
necessarily $\varphi_0=0$ because $U\Omega$ is unitary, hence
$\varphi_1 \in \cV_1$. That $\cV_1$ is the kernel implies that
$\dim \cV_1 + \dim (I - F_1 U \Omega) \cF_1 = \dim \cF_1 $.\footnote{Here the
  assumption of finite dimension of the spaces is invoked.} As a
consequence\footnote{Note that in general
  $(I - F_1 U \Omega): \cF_1 \to \cF_1$ is not a normal operator.}
\be
\cF_1  =  \cV_1  \oplus (I - F_1 U \Omega) \cF_1
.
\label{eq:1.6}
\ee
To show this, first note that $\cV_1 \perp (I - F_1 U \Omega)\cF_1$, hence
$(I - F_1 U \Omega)\cF_1 \subseteq \cF_1 \ominus \cV_1$. But the spaces
$(I - F_1 U \Omega)\cF_1$ and $\cF_1 \ominus \cV_1$ have the same dimension
so they must coincide.

Next we make use of the relation
\be
F_1 U \cH_0 \oplus \cF_1 \cap U \cH_1 = \cF_1
\,,
\label{eq:1.20}
\ee
which is a consequence of Lemma \ref{lm:3} and $\cF = U \cH_0 \oplus U
\cH_1$. Then, using $\cH_1 = \Omega \cF_1$,
\bes & \cV_1 \subseteq\cF_1 \cap U \Omega \cF_1 = \cF_1 \cap U\cH_1 =
\cF_1 \ominus F_1 U \cH_0
\\
&
\implies ~ F_1 U \cH_0 \subseteq \cF_1 \ominus \cV_1
= (I - F_1 U \Omega)\cF_1 .
\label{eq:1.8}
\ees
This guarantees that \nec{1.3b} has solutions for arbitrary $\psi_0$.

The solutions to \nec{1.3b} have the form $\phi_1 = \phi_1' + \varphi_1$,
where $\varphi_1$ is an arbitrary vector of $\cV_1$ while
$\phi_1' \in \cF_1\ominus\cV_1$ is unique. In the first \eq{1.3}
$F_0 U \Omega \varphi_1 = F_0 \varphi_1 = 0$, therefore
\be
\phi_0 = F_0 U \psi_0 + F_0 U \Omega \phi_1'
\label{eq:1.9}
\ee
uniquely determines $\phi_0$. This defines a linear map $\psi_0 \to \phi_0$.
$U$ is unitary hence
$\|\psi_0\|^2 + \|\psi_1\|^2 = \|\phi_0\|^2 + \|\phi_1\|^2 $, and $\Omega$ is
also unitary, then $\|\psi_1\| = \|\phi_1\|$. This implies
$\|\psi_0\| = \|\phi_0\|$. \qed

\newdefinition{df:1} ~{\em (Reduction of a unitary map by a unitary
  suboperator\footnote{More precisely, the suboperator is $\Omega^{-1}$. More
    properly $\Omega$ could be called an anti-suboperator of $U$.})}
~The map $\psi_0\to\phi_0$ obtained in Theorem \ref{th:1} will be denoted
$\OcU$.  The unitary operator $\OcU:\cH_0\to\cF_0$ is the {\em reduction
  of\/ $U:\cH\to\cF$ to $\cH_0\to\cF_0$ with connection
  $\Omega:\cF_1\to\cH_1$}, or simply, the reduction of $U$ to $\cH_0$. \ignore{ The
subspaces $\cH_1$ and $\cF_1$ have been {\em reduced}. }  \fin

\begin{figure}[ht]
  \begin{center}
    \includegraphics[height=40mm]{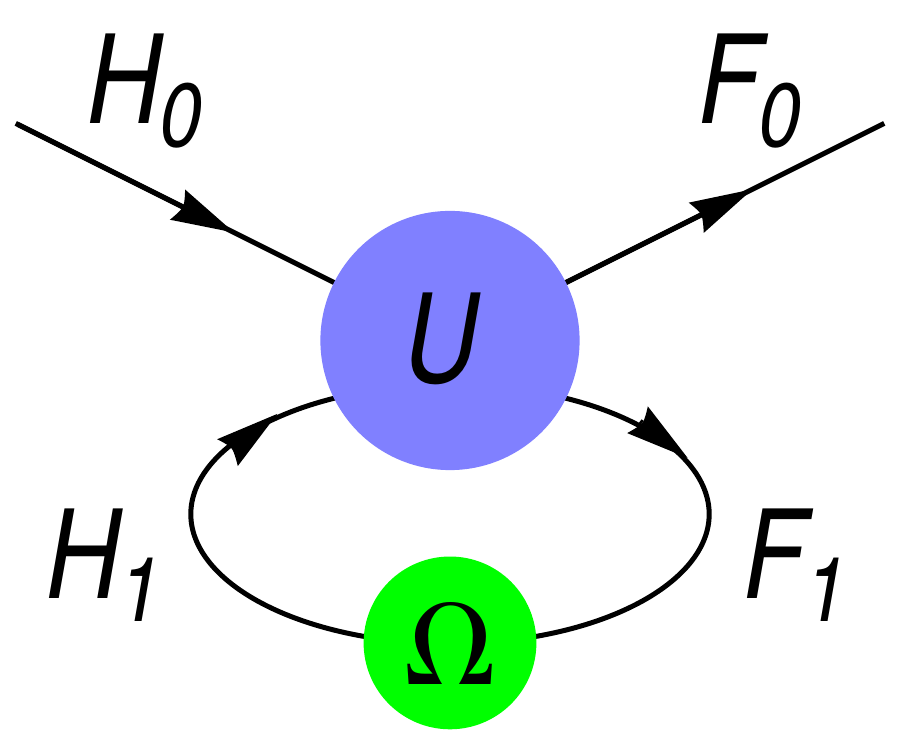}
    \end{center}
\caption{Schematics of the reduction.}
\label{fig:1}
\end{figure}

Fig. \ref{fig:1} shows a cartoon of the unitary reduction setting.

\newparticularcase{pc:1}
~A particular case corresponds to $\cH = \cF$, with $\cH_0 = \cF_0$ and
$\cH_1 = \cF_1$, and $\Omega$ is the identity operator in $\cH_1$. This can
be denoted $I_{\cH_1}\cdot U$ or even $\cH_1 \cdot U$, the reduction of $U$
to the subspace $\cH \ominus \cH_1$. \fin

The component of $\phi_1$ along the subspace $\cV_1$ is left undetermined and
it can be set to $0$ as it does not have a contribution to the final value of
$\phi_0$. Such component does not originate from the initial $\psi_0$, and the
pair of subspaces $\cV_1\subseteq \cF_1$ and its image
$\Omega \cV_1\subseteq \cH_1$ completely decouple from the rest in the map
$\OcU$.

\newlemma{lm:1} ~Let $\cV\subseteq \cF_1 \cap U \cH_1$ be the largest subspace
which is invariant under $U\Omega: \cF_1 \to U \cH_1$, namely the sum of all
$U\Omega$-invariant subspaces. In particular $\cV_1\subseteq \cV$. Then the map
$\OcU:\cH_0\to\cF_0$ is unchanged if the subspaces $\cV$ and $\Omega \cV$ are
stripped from $\cF_1$ and $\cH_1$ respectively (and so from $\cF$ and
$\cH$). Also unchanged are the operators
\be
(U\Omega F_1)^n UH_0 \qquad n\ge 0
\label{eq:1.5}
.
\ee

\proof  ~$\OcU$ will be unchanged if the equations \nec{1.1} still
have a solution adding the new condition $\varphi_V=0$, where
$\varphi_V := F_V \phi_1$ and $F_V$ is the orthogonal projector onto $\cV$.

Let $\cF_1' := \cF_1 \ominus \cV$. Then the vector $\phi_1$ can be split as
$\phi_1 = \phi_1' + \varphi_V$, where $\phi_1'\in\cF_1'$ and
$\varphi_V \in \cV$. The projection of the equation
$\phi_0 + \phi_1 = U( \psi_0 + \Omega \phi_1 )$ onto $\cV$ produces
\be
\varphi_V
=
F_V U \psi_0 + F_V U\Omega \phi_1' + F_V U\Omega \varphi_V
.
\ee
In the RHS $F_V U \psi_0=0$ because $\cV \subseteq U \cH_1 \perp U \cH_0$.  Also
$F_V U\Omega \phi_1' = 0$ because $\cV \perp \cF_1'$ hence
$\cV=U\Omega \cV \perp U\Omega \cF_1'$. In addition
$F_V U\Omega \varphi_V = U\Omega \varphi_V$ due to $\cV=U\Omega \cV$. The
projected equation becomes $\varphi_V = U\Omega \varphi_V$, and this implies
that $\varphi_V = \varphi_1 \in \cV_1$. As already noted $\varphi_1$ is not
determined by the equations and can be chosen to be zero. Therefore the constraint
$\varphi_V = 0$ does not modify $\phi_0$ nor the map $\OcU$.

To prove that the operators in \eq{1.5} are unchanged, let $F_1'$ be the
projector onto $\cF_1'$. The following relations are verified,
\be
F_V U H_0 = F_V F_1' = 0,
\qquad
[F_V,U\Omega] = 0
,
\qquad
F_1 = F_1' + F_V
.
\ee
The statement is that $F_1'$ can be used instead of $F_1$ in \nec{1.5}.
This is trivial for $n=0$. For $n=1$
\bes
 U \Omega F_1 U H_0 &=
 U \Omega ( F_1' + F_V ) U H_0
 \\
 &=  U \Omega F_1' U H_0 
 .
 \ees
By induction, for $n \ge 2$,
\bes
( U \Omega F_1)^n U H_0 &=
U \Omega (F_1' + F_V) (U \Omega F_1')^{n-1} U H_0
\\ &=
(U \Omega F_1')^n U H_0
\ees
using $F_V U\Omega F_1' = U \Omega F_V F_1'=0$.
\qed

The lemma implies that in the construction of $\OcU$, the subspaces $\cV$ and
$\Omega \cV$ completely decouple from the rest.

\newparticularcase{pc:2}
~An extreme instance of decoupling takes place
when $\cF_0 = U\cH_0$ and $\cF_1 = U\cH_1$. In this case
$\cV = \cF_1$ and the reduction $\OcU$ is just the restriction of $U$ to
$\cH_0 \to \cF_0$, regardless of the choice of $\Omega$. \fin

\newparticularcase{pc:3} ~An opposite extreme situation is $\cF_1 = U\cH_0$
and $\cF_0 = U\cH_1$. In this case $\cV=0$ and $\OcU = U \Omega U$ as an
operator from $\cH_0$ to $\cF_0$. \fin

The map $U\Omega : \cV \to \cV$ is unitary, hence clearly
$\| F_1 U \Omega \| = 1$ is a necessary (but not sufficient) condition for a
non vanishing $\cV$ or $\cV_1$. The following lemma provides a necessary and
sufficient condition.

\newlemma{lm:2} ~Let $\cH$ be a Hilbert space, $U:\cH\to\cH$ a unitary
operator and $\cW$ a non-zero subspace of $\cH$. Then $\cW$ contains a
non-zero $U$-invariant subspace if and only if $\| (W U W)^d \|=1$, where $W$
is the orthogonal projector onto $\cW$ and $d=\dim \cW$.

\proof ~If there is a non-zero invariant subspace in $\cW$, then there is an
eigenvector of $U$ in $\cW$ and this implies $\| (W U W)^d \|=1$.  Let us now
assume that $\| (W U W)^d\|=1$. In this case there is a unit vector
$u_0\in \cH$ such that $u_d:=W (U W)^d u_0$ is also a unit
vector.\footnote{Once again invoking that the spaces are finite-dimensional.}
Clearly $u_0\in \cW$, otherwise $\|W u_0\|<1$ and this would imply
$\|u_d\|<1$.  By the same token, all the vectors $u_k:=U^k u_0$ for
$k=0,\ldots,d$ must fulfill $u_k\in \cW$ to ensure $\|u_d\|=1$. Since no $d+1$
vectors can be linearly independent in $\cW$ it follows that one of them,
$u_{k_0}$, must be a linear combination of the previous ones, and in this case
the vectors $u_0,\ldots,u_{k_0-1}$ span a non-zero subspace which is invariant
under $U$.  \qed

The same statement holds for any exponent $d'\ge d$ instead of $d$.

\newexample{ex:1} ~As a simple example, let\footnote{The asterisk denotes
  complex conjugation.}
\be
U = \PM{ \malpha & - \omega \mbeta^* \\
  \mbeta & \omega \malpha^* },
\quad
\malpha,\mbeta,\omega \in \C, \quad
|\malpha|^2 + |\mbeta|^2 = |\omega| = 1
\,.
\ee
$U$ is a generic unitary matrix in $\cH=\cF=\C^2$.  Furthermore
$\cH_0 = \cF_0$ and $\cH_1 = \cF_1$ are spanned by the upper and lower
components, respectively. The map $\Omega$ is codified by a number
$\Omega\in\C$, $|\Omega|=1$. The equations defining the reduction are
\be
\PM { \phi_0 \\ \phi_1 } = \PM{ \malpha & - \omega \mbeta^* \\
  \mbeta & \omega \malpha^* }
\PM { \psi_0 \\ \psi_1 }
,
\qquad
\psi_1 = \Omega \phi_1 
\,.
\ee

Three cases can be distinguished:
\begin{itemize}

\item[i)] $\mbeta \neq 0$. Then $\cF_1 \cap U\cH_1 = 0$.

\item[ii)] $\mbeta=0$ and $\malpha \neq \omega\Omega$. Then $\cV=U\cH_1=\cF_1$
  and $\cV_1=0$.

  In any of these two cases $\phi_1$ is unique, and
\be
\phi_1 = \frac{\mbeta }{ 1-\omega \Omega \malpha^* } \psi_0,
\qquad
\phi_0 =
- \frac{1}{\omega \Omega} \frac{\malpha - \omega \Omega }{(\malpha - \omega
  \Omega)^*} \psi_0
.
\ee
The map $\psi_0 \to \phi_0$ depends only on the combination $\omega \Omega$
and not on (the phase of) $\mbeta$. It is also noteworthy that $\|\phi_1\|$ is
not bounded, for fixed $\|\psi_0\|$.

\item[iii)] $\malpha = \omega\Omega$. Then $\cV_1=\cF_1$, $\phi_1$ is
  arbitrary, and $\phi_0 = \malpha \psi_0$.
  
\end{itemize}

In all cases the map $\psi_0 \to \phi_0$ is manifestly unitary. The map $\OcU$
is not a continuous function of $U$ and $\Omega$ at
$\malpha=\omega\Omega$. The value of $\OcU$ at $\malpha=\omega\Omega$
coincides with the limit taken on the set $|\malpha|=1$. On this set
$\mbeta=0$ and $\phi_1$ vanishes (for $\malpha \neq \omega\Omega$).  In the
cases i) and iii) reciprocity applies (see Theorem \ref{th:2} below).  \fin

\subsection{  Some properties of the reduction }

The following proposition shows that performing two reductions sequentially
is equivalent to doing the reduction in a single step.

\newproposition{pr:1} ~{\em (Sequential reductions)}
~Let $\cH=\cH_0\oplus\cH_1\oplus\cH_2$ and $\cF=\cF_0\oplus\cF_1\oplus\cF_2$
be Hilbert spaces, and let $U:\cH \to \cF$, $\Omega_1: \cF_1\to \cH_1$ and
$\Omega_2: \cF_2\to \cH_2$ be unitary maps. Then
\be
(\Omega_1\oplus\Omega_2) \cdot U =
\Omega_1 \cdot (\Omega_2 \cdot U)
\ee

\proof ~Let us make use of the decompositions
$\psi = \psi_{01} + \psi_2 = \psi_0 + \psi_1 + \psi_2$, and
$\phi = \phi_{01} + \phi_2 = \phi_0 + \phi_1 + \phi_2$.  The first reduction
$\phi_{01} = \Omega_2\cdot U \psi_{01}$ is determined by the equations
\be
\phi = U \psi
,
\qquad
\Omega_2\phi_2 = \psi_2
,
\ee
upon elimination of $\psi_2$ and $\phi_2$. Likewise, the second
reduction is determined by the equations
\be
\phi_{01} = \Omega_2\cdot U \psi_{01}
,
\qquad
\Omega_1\phi_1 = \psi_1
,
\ee
upon elimination of $\psi_1$ and $\phi_1$. Hence the final map
$\psi_0 \to \phi_0$ is determined by the conditions
\be
\phi = U \psi
,
\qquad
\Omega_1 \phi_1 = \psi_1
,
\qquad
\Omega_2 \phi_2 = \psi_2
.
\ee
This is the same set of equations that would be produced by eliminating
$\cH_1 \oplus \cH_2$ with the connection $\Omega_1 \oplus \Omega_2$.
\qed

Another similar proposition is as follows:

\newproposition{pr:2}
~Let $\cH=\cH_0\oplus\cH_1$ and $\cF=\cF_0\oplus\cF_1$ be Hilbert spaces, and
let $U:\cH \to \cF$ and $\Omega: \cF_1\to \cH_1$ be unitary maps. Then
\be
\Omega^{-1} \cdot U^{-1} = ( \Omega \cdot U )^{-1} 
\ee

\proof ~Let us make use of the decompositions
$\psi= \psi_0 + \psi_1 $, and
$\phi = \phi_0 + \phi_1 $.  The reduction
$\phi_0 = \Omega\cdot U \psi_0$ is determined by the equations
\be
\phi = U \psi
,
\qquad
\Omega \phi_1 = \psi_1
,
\ee
upon elimination of $\psi_1$ and $\phi_1$. These conditions are equivalent to
\be
\psi = U^{-1} \phi
,
\qquad
\Omega^{-1} \psi_1 = \phi_1
,
\ee
which define the map $\psi_0 = \Omega^{-1} \cdot U^{-1} \phi_0$.
\qed

Some properties under chiral transformations are also easily established:

\newproposition{pr:3} ~Let $Q_0$, $Q_1$, $R_0$ and $R_1$ be unitary maps
acting in $\cH_0$, $\cH_1$, $\cF_0$ and $\cF_1$, respectively, and they are
naturally extended as the identity in the complementary spaces.
Then
\be
\Omega \cdot ( R_0 R_1 U Q_0 Q_1 )
= R_0 ( Q_1 \Omega R_1 \cdot U ) Q_0
.
\label{eq:1.22}
\ee

\proof ~
The map $\Omega \cdot ( R_0 R_1 U Q_0 Q_1 )$ is defined by the following
equations
\be
\phi_0 + \phi_1  = R_0 R_1 U Q_0 Q_1 (\psi_0 + \psi_1)
,
\qquad \psi_1 = \Omega \phi_1
.
\ee
This implies
\be
R_0^{-1} \phi_0 + R_1^{-1} \phi_1  =  U (Q_0 \psi_0 + Q_1 \Omega \phi_1 )
.
\ee
Let us define
\be
\phi_0' = R_0^{-1} \phi_0
,
\qquad
\phi_1' = R_1^{-1}\phi_1
,
\qquad
\psi_0' = Q_0 \psi_0
.
\ee
The equation is then equivalent to
\be
\phi_0' + \phi_1'  =  U (\psi_0' + Q_1 \Omega R_1 \phi_1' )
.
\ee
Therefore $( Q_1 \Omega R_1 \cdot U) \psi_0' = \phi_0'$, and $\nec{1.22}$
follows.
\qed

As well as properties under tensor product:

\newproposition{pr:5} ~Let $U: \cH_0\oplus\cH_1 \to \cF_0\oplus\cF_1$,
$\Omega : \cF_1 \to \cH_1$,
and $V : \cH' \to \cF'$, be unitary operators. Then
\be
( \Omega \otimes V^{-1}) \cdot ( U \otimes V ) = ( \Omega \cdot U ) \otimes V
\ee

\proof
~The map on the LHS is defined by the equations
\be
\Phi_0 + \Phi_1 =  U \otimes V ( \Psi_0 + \Psi_1 ),
\qquad
\Omega \otimes V^{-1} \Phi_1  =\Psi_1
,
\label{eq:1.27}
\ee
where $\Psi_0$, $\Psi_1$, $\Phi_0$ and $\Phi_1$ are vectors in
$\cH_0 \otimes \cH'$, $\cH_1 \otimes \cH'$, $\cF_0 \otimes \cF'$ and
$\cF_1 \otimes \cF'$, respectively. The map is $\Psi_0 \to \Phi_0$ and it is
sufficient to consider the case of a separable $\Psi_0$,
\be
\Psi_0 = \psi_0 \otimes \psi',
\qquad \psi_0 \in \cH_0, \quad \psi' \in \cH'
.
\ee
Since $\Phi_0$ is uniquely defined by the equations, new constraints on $\Phi$
and $\Psi_1$ can be added if they prove to be compatible. So we add the
equations
\bes
\Phi_0 &= \phi_0 \otimes \phi'
,\quad
\Phi_1 = \phi_1 \otimes \phi'
,\quad
\\
\Psi_1 &= \psi_1 \otimes \psi'
,\quad
\phi' = V \psi'
,\\
\phi_0 & \in \cF_0, \quad
\phi_1 \in \cF_1, \quad
\psi_1 \in \cH_1, \quad
\phi' \in \cF' 
\,.
\ees
Substitution in \nec{1.27} yields
\bes
(\phi_0 + \phi_1 ) \otimes \phi' &= U(\psi_0 + \psi_1) \otimes V \psi'
,
\\
\Omega \phi \otimes V^{-1} \phi' &= \psi_1 \otimes \psi'
.
\ees
These equations admit the solution $\phi_0 = (\Omega \cdot U)\psi_0$, hence
$\Phi_0 = (\Omega \cdot U) \otimes V \psi_0 \otimes \psi'$.
\qed

Until now we have considered the elimination (reduction) of one of the two
subspaces in $\cH = \cH_0 \oplus \cH_1$. One can investigate the relation
between the two reductions obtained by eliminating either $\cH_1$ or
$\cH_0$. Specifically, under what conditions the map
$\Omega^{-1}:\cH_1\to\cF_1$ is obtained as the reduction when
$(\Omega\cdot U)^{-1}:\cF_0\to \cH_0$ is used as the connection:

\newtheorem{th:2} ~{\em (Reciprocity)} ~ Let
$U:\cH_0 \oplus \cH_1 \to \cF_0 \oplus \cF_1$ and $\Omega:\cF_1 \to \cH_1$ be
unitary maps between Hilbert spaces, with $\cV_1$ and $F_1$ as defined
previously. Then the following assertions are equivalent
\be
(\Omega\cdot U)^{-1} \cdot U = \Omega^{-1}
,
\label{eq:1.22a}
\ee
\be
\cV_1 = \cF_1 \cap U \cH_1
.
\label{eq:1.22b}
\ee

\proof ~
The equations defining the map $(\Omega\cdot U)^{-1} \cdot U $
are
\be
\phi_0 + \phi_1 = U ( \psi_0 + \psi_1),
\qquad
\psi_0 = (\Omega\cdot U)^{-1} \phi_0
\,,
\label{eq:1.25}
\ee
with $\psi_0\in\cH_0$, $\psi_1\in\cH_1$, $\phi_0\in\cF_0$ and
$\phi_1\in\cF_1$.  The map is $\psi_1 \to \phi_1$ and it is invertible. Theorem
\ref{th:1} then guarantees that for every $\phi_1$ there is a solution for
$(\psi_0, \psi_1,\phi_0)$ and furthermore $\psi_1$ is unique.  Hence, the
condition \nec{1.22a} requiring $\psi_1 = \Omega \phi_1$ is equivalent to
\be
\forall \phi_1 \quad
\exists \psi_0, \phi_0
\quad
\phi_0 + \phi_1 = U ( \psi_0 + \Omega\phi_1),
\quad 
\psi_0 = (\Omega\cdot U)^{-1} \phi_0
\,.
\ee
The second equation amounts to
$\exists \phi_1' \in \cF_1 ~~ \phi_0 + \phi_1' = U ( \psi_0 + \Omega\phi_1')$,
which is a consequence of the first one and it can be dropped. The remaining
equation is equivalent to the pair
\bes
\forall \phi_1 ~
\exists \psi_0, \phi_0
\quad
\phi_0 &= F_0 U \psi_0 + F_0 U \Omega \phi_1
,
\quad
\\
\phi_1 &= F_1 U \psi_0 + F_1 U \Omega \phi_1
.
\ees
The first equation just provides $\phi_0$ in terms of $\phi_1$ and $\psi_0$.
The only nontrivial condition is the second one, which is equivalent to
\be
(I-F_1U \Omega) \cF_1 \subseteq F_1 U \cH_0
\,.
\label{eq:1.22c}
\ee
Thus the conditions \nec{1.22a} and \nec{1.22c} are equivalent.

On the other hand, according to \nec{1.8},
$ F_1 U \cH_0 \subseteq (I- F_1U\Omega) \cF_1 = \cF_1 \ominus \cV_1$
always. Hence, \nec{1.22c} is equivalent to
$ F_1 U \cH_0 = \cF_1 \ominus \cV_1$. In view of \nec{1.20} this is equivalent
to \nec{1.22b}.  \qed

\newcorollary{cor:1}
~$\cV_1 = \cF_1 \cap U \cH_1 \implies \cV_1' = \cF_0 \cap U \cH_0$. Here
$\cV_1' := \{ \phi_0 \in \cF_0 ~|~ U \Omega' \phi_0 = \phi_0 \}$, and
$\Omega' := (\OcU)^{-1}$.

\proof ~ According to Theorem \ref{th:2}, $\cV_1 = \cF_1 \cap U \cH_1 $
implies $\Omega' \cdot U = \Omega^{-1}$. In turn,
\be
(\Omega' \cdot U)^{-1} \cdot U = \OcU = \Omega'{}^{-1}
.
\ee
Then Theorem \ref{th:2} requires $\cV_1' = \cF_0 \cap U \cH_0$. \qed

The implication in the other direction does not hold. 

$\cF_1 \cap U \cH_1$ does not depend on $\Omega$ and
$\cV_1 \subseteq \cF_1 \cap U \cH_1$ always.  Theorem \ref{th:2} implies that
to have reciprocity between $\Omega$ and $\OcU$ the eigenspace $\cV_1$ must
completely fill the available space $\cF_1 \cap U \cH_1$, and in particular
$\cV = \cV_1$.

A sufficient condition for \nec{1.22b} to hold is $U \cH_1 \subseteq \cF_0$,
since in this case $\cV_1= \cF_1 \cap U \cH_1 = 0$. This condition does not
depend on $\Omega$. For this kind of $U$ and decompositions $\cH_0\oplus\cH_1$
and $\cF_0\oplus\cF_1$, any operator $U_1:\cH_1 \to \cF_1$ can be obtained as a
reduction $\Omega_0\cdot U$ taking a suitable connection $\Omega_0$, in
particular choosing $\Omega_0 = (U_1^{-1} \cdot U)^{-1}$. Of course this is
possible because $U \cH_1 \subseteq \cF_0 $ guarantees
$\dim \cH_0 \ge \dim \cH_1$, and so there enough freedom in the $0$ sector to
reproduce any $U_1$.

In the aforementioned Particular case \ref{pc:2}, where $U\cH_0=\cF_0$ and
$U\cH_1=\cF_1$, $\OcU = U|_{\cH_0}$ regardless of $\Omega$, so in general
$\Omega^{-1}$ will not be recovered as reduction using $(\OcU)^{-1}$ as
connection. In fact, any connection $\Omega':\cF_0 \to \cH_0$ produces
$ \Omega' \cdot U = U|_{\cH_1}$. This will be $\Omega^{-1}$ if and only if
$U \Omega = I_{\cH_1}$, i.e., $\cV_1 = \cF_1$, in agreement with Theorem
\ref{th:2}.

In the Particular case \ref{pc:3} ~$U \cH_0 = \cF_1$ and $U \cH_1 = \cF_0 $,
$\cV = \cV_1= 0$ and also $\cF_1 \cap U \cH_1=0$; the premise in Theorem
\ref{th:2} is satisfied and there is reciprocity between connections and
reductions. Indeed, for any connection $\Omega \cdot U = U \Omega U$ (acting
on the appropriate space) then
\be
( \Omega \cdot U )^ {-1} \cdot U =
( U \Omega U )^{-1} \cdot U =
U ( U \Omega U )^{-1} U = \Omega^{-1}
\,.
\ee
The property $\Omega \cdot U = U \Omega U$ for arbitrary $\Omega$ implies that
for this type of $U$ one can choose the connection $\Omega$ to produce any
given operator $U_0:\cH_0\to\cF_0$ through the reduction
$U_0 = \Omega \cdot U$.

\subsection{  Alternative forms of the reduction }

Combining \eqs{1.3b} and \nec{1.9} from the proof of Theorem \ref{th:1},
it follows that the reduced map can be expressed as
\be
\OcU =
F_0 U H_0 +
F_0 U \Omega F_1
(I - F_1 U \Omega F_1 )^{-1}
F_1 U H_0
,
\label{eq:1.10a}
\ee
understood as an operator $\cH_0\to\cF_0$.  The operator
$I - F_1 U \Omega F_1$ does have an inverse within $\cF \ominus \cV$ which
is the only range attainable by $F_1U\cH_0$. Introducing the notation
\bes
U_{00} &:= F_0 U H_0,
\quad
\tU_{01} := F_0 U\Omega F_1,
\quad
U_{10} := F_1 U H_0,
\quad
\\
\tU_{11} &:= F_1 U \Omega F_1
,
\label{eq:2.44}
\ees
\eq{1.10a} can be expressed as
\be
\OcU =
U_{00} + \tU_{01} (I - \tU_{11} )^{-1} U_{10}
.
\label{eq:1.10}
\ee

The reduction can also be introduced as the sum of a series:

\newtheorem{th:3}
~The following series is absolutely convergent and its sum is $\OcU$.
\be
\OcU = U_{00} + \tU_{01}( I + \tU_{11} + \tU_{11}^2+ \tU_{11}^3+ \cdots )U_{10}
.
\label{eq:1.1c}
\ee

\proof ~Clearly when the series is absolutely
convergent its sum is given by \nec{1.10}.  Let us show that the series in
\eq{1.1c} is indeed absolutely convergent.

~From Lemma \ref{lm:1}, each of the terms of the series is unchanged if the
subspaces $\cV$ and $\Omega \cV$ are removed from $\cF$ and $\cH$,
respectively. Therefore, without loss of generality we can assume that the map
$U\Omega$ is defined in $\cF_1 \ominus \cV$, and has no non-zero invariant
subspaces.

The operators $U_{j0}$ and $\tU_{j1}$ , for $j\in\{0,1\}$, are subunitary,
hence $\| (U_{j0})^n \| \le 1$ and $\| (\tU_{j1})^n \| \le 1$, for $n\in\Z$
and $n\ge 0$.  Let $d:=\dim \cF_1 \ominus \cV$ and $N:=\| (\tU_{11})^d \|$,
then $N \le 1$.

When $\tU_{11} = 0$ the series in \nec{1.1c} terminates. Let us assume that
$\tU_{11} \neq 0$, hence $d \ge 1$, and for a generic term of the series
$\| \tU_{01} (\tU_{11})^{ kd+r} U_{10} \| \le N^k$. Since $\cF_1\ominus \cV$
contains no non-zero $U\Omega$-invariant subspace, Lemma \ref{lm:2} implies
that $N < 1$. Then each of the subseries for $r=0,\ldots,d-1$ is absolutely
convergent. \qed

Note that $\|(\tU_{11})^k\| = 1$ for some $1 \le k<d$ is not excluded.

The parameter $N$ is a relevant one in the reduction since it
controls the convergence of the series. Such convergence is by no means
uniform in the set of unitary operators $U$ and $\Omega$ since $N$ can be
arbitrarily close to unity. As shown in the Example \ref{ex:1} in $\C^2$, the map
$(U,\Omega) \to \OcU$ is not continuous for some values of $(U,\Omega)$.

Two further alternative forms of the reduction are as follows:

\newproposition{pr:4} The map $\OcU$ can be expressed as
\be
\OcU = F_0 ( I - U \Omega F_1 )^{-1} U H_0 
\,,
\label{eq:1.37}
\ee
and also as
\be
\OcU = \lim_{ n \to \infty } (F_0 + U \Omega F_1 )^n U H_0
\,.
\label{eq:1.38a}
\ee

\proof
\be
\phi = U ( \psi_0 + \psi_1 ), \quad
\psi_1 = \Omega \phi_1 
~\implies~
\phi = U ( \psi_0 + \Omega F_1 \phi )
.
\ee
As already shown, $\phi$ has a unique solution within $\cF \ominus \cV$, and
$ U \psi_0 \in \cF \ominus \cV$, hence
$\phi = ( I - U \Omega F_1 )^{-1} U \psi_0 $ and so
\be
\phi_0 =  F_0 ( I - U \Omega F_1 )^{-1} U \psi_0 
\ee
which is \nec{1.37}. Alternatively, an expansion of
$F_0 ( I - U \Omega F_1 )^{-1} U \psi_0$ as a geometrical series reproduces
the series in \nec{1.1c}, which is absolutely convergent. The expansion is
justified since, within $\cF \ominus \cV$, either $\tU_{11} = 0$ or
$\|\tU_{11}^d\| < 1$, where $d = \dim\cF_1\ominus\cV$ and $d \ge 1$.

For the second relation, \eq{1.38a}, it is easily verified that
\be
(F_0 + U \Omega F_1 )^n =
F_1 (U\Omega F_1)^n +  \sum_{k=0}^n F_0 (U\Omega F_1)^k
,\quad n=0,1,2,\ldots
\label{eq:1.40}
\ee
This is to be multiplied by $U H_0$ on the right. By the previous convergence
arguments, the first term in the RHS goes to $0$ when $n\to\infty$, while the
second term reproduces precisely the same series a
$F_0 ( I - U \Omega F_1 )^{-1} U H_0$.  \qed

\subsection{ Relation to Kraus operators }

The relation \nec{1.38a} is particularly illuminating as regards to the
interpretation of the reduction. In the language of quantum mechanics,
$\psi_0$ represents a state vector in $\cH_0$ reaching a gate $U$ to yield the
outgoing state $U\psi_0$. The operator $F_0 + U \Omega F_1 $ lets pass the
component along $\cF_0$ while that along $\cF_1$ is sent through the gate
$\Omega$ and enters $U$ again, to repeat the process. Eventually the $\cF_1$
component fades away and all the flux goes into $\cF_0$ as the state
$\phi_0$. The parameter $N$ would control the average number of passes
through $\Omega$.

\Eq{1.38a} encodes in compact form an iterative solution of \eqs{1.1}. More explicitly,
\bes
\psi_0 \to U\psi_0 &= \phi_0^{(1)} + \phi_1^{(1)}
,
\qquad
\\
U \Omega \phi_1^{(n)} &= \phi_0^{(n+1)} + \phi_1^{(n+1)} \quad
\forall n \ge 1
,
\\
\phi_0 &= \sum_{n=1}^\infty \phi_0^{(n)}
,\qquad
\phi_1 = \sum_{n=1}^\infty \phi_1^{(n)} 
.
\label{eq:1.53}
\ees
The expansion
\be
\phi_0 = \phi_0^{(1)} + \phi_0^{(2)} + \cdots + \phi_0^{(n)} + \cdots 
\ee
is the series in \nec{1.1c}. The terms are classified by the number of
times the operator $U$ acts. Each term is related to $\psi_0$ as
\be
\phi_0^{(n)} = A_n \psi_0,
\qquad
A_n := F_0 (U\Omega F_1)^{n-1} U H_0 \quad \forall n \ge 1,
\ee
and by construction
\be
\Omega \cdot U = \sum_{n=1}^\infty A_n
\label{eq:1.55}
.
\ee
What is remarkable is the following result:

\newproposition{pr:10} The  $\{ A_n \} $ are Kraus operators, that
is,\footnote{The symbol ${}^\dagger$ denotes Hermitian adjoint.}
\be
\sum_{n=1}^\infty A_n^\dagger A_n = I_{\cH_0}. 
\label{eq:1.56}
\ee
($I_{\cH_0}$ denotes the identity operator in $\cH_0$.)

\proof ~$\|\psi_0\|^2 = \|\phi_0^{(1)}\|^2 + \|\phi_1^{(1)}\|^2$ because $U$
is unitary and also
$\|\phi_1^{(n)}\|^2 = \|\phi_0^{(n+1)}\|^2 + \|\phi_1^{(n+1)}\|^2$ for $n\ge1$
because $U\Omega$ is unitary, hence
\be
\|\psi_0\|^2 = \sum_{k=1}^n \| \phi_0^{(k)} \|^2 +  \| \phi_1^{(n)} \|^2
\quad \forall n \ge 1
.
\ee
In addition $\|\phi_1^{(n)}\| \to 0$ for large $n$. As a consequence
\be
\|\psi_0\|^2 = \sum_{n=1}^\infty \| \phi_0^{(n)} \|^2
=
\sum_{n=1}^\infty \| A_n \psi_0 \|^2
.
\ee
Since this holds for all $\psi_0$ in a complex Hilbert space \nec{1.56}
follows. \qed

This implies the relation
\be
\| \sum_{n=1}^\infty  \phi_0^{(n)} \|^2
= \sum_{n=1}^\infty \| \phi_0^{(n)} \|^2
,
\ee
even though the various terms are not orthogonal.

The Kraus operators define a quantum channel
\be
T_{U,\Omega}(\rho) := \sum_{n=1}^\infty  A_n \rho A_n^\dagger
,
\label{eq:2.59}
\ee
where $\rho$ is a density matrix operator acting in $\cH_0$. They also define
a positive operator-valued measure (POVM)
\be
\Pi_n := A_n^\dagger A_n,
\qquad
\Pi_n\ge 0, \qquad
\sum_{n=1}^\infty \Pi_n = I_{\cH_0}
.
\ee
In quantum mechanics the observable $n$ would represent the number of passes
through $U$ before irreversibly going to the $\cF_0$ sector, and
$\esp{\psi_0|\Pi_n|\psi_0}$ the probability of such $n$.  The quantum channel
provides the mixed state left after a non-selective measurement of the
observable $n$.

In the context of quantum channels the Kraus operators act incoherently, what
is remarkable here is that, according to \nec{1.55}, the coherent sum of the
$A_n$ produces a unitary operator.

In general the quantum channel $T_{U,\Omega}$ introduced above is not a
unitary quantum channel, however it enjoys the following property

\newproposition{pr:11} The quantum channel $T_{U,\Omega}$ is unital.

\proof ~Unital means that $T_{U,\Omega}(I_{\cH_0}) = I_{\cF_0}$. This is
equivalent to
\be
\sum_{n=1}^\infty A_n A_n^\dagger = I_{\cF_0}
,
\ee
and also equivalent to the statement that the operators $\{ A_n^\dagger \}$
are also Kraus operators. In fact, the operators $A_n$ can be expressed as
\be
A_n = F_0 (U H_1 \Omega F_1)^{n-1} U H_0 
\ee
(explicitly inserting the redundant projector $H_1$) and then for its
adjoint operator
\bes
A_n^\dagger &= H_0 U^{-1} ( F_1 \Omega^{-1} H_1 U^{-1} )^{n-1} F_0
\\
&= H_0 (U^{-1} F_1 \Omega^{-1} H_1)^{n-1} U^{-1} F_0
.
\ees
These are precisely the Kraus operators corresponding to the reduction
$\Omega^{-1}\cdot U^{-1}$.  \qed

Being unital is a necessary condition (but not sufficient for $\dim\cH_0>2$
\cite{Kummerer:1987,Audenaert:2008}) for $T_{U,\Omega}$ to be a convex
combination of unitary channels (a random unitary channel) that is, of the
form
\be
\rho \to  \sum_j p_j U_j \rho U_j^{-1},
\quad
p_j\ge0, \quad
\sum_j p_j=1, \quad
U_j^\dagger U_j = I_{\cH_0}
.
\ee
The question whether $T_{U,\Omega}$ is necessarily a random unitary channel or
not is not pursued in the present work.

\subsection{ Unitary networks }

\begin{figure}
  \begin{center}
    \includegraphics[height=60mm]{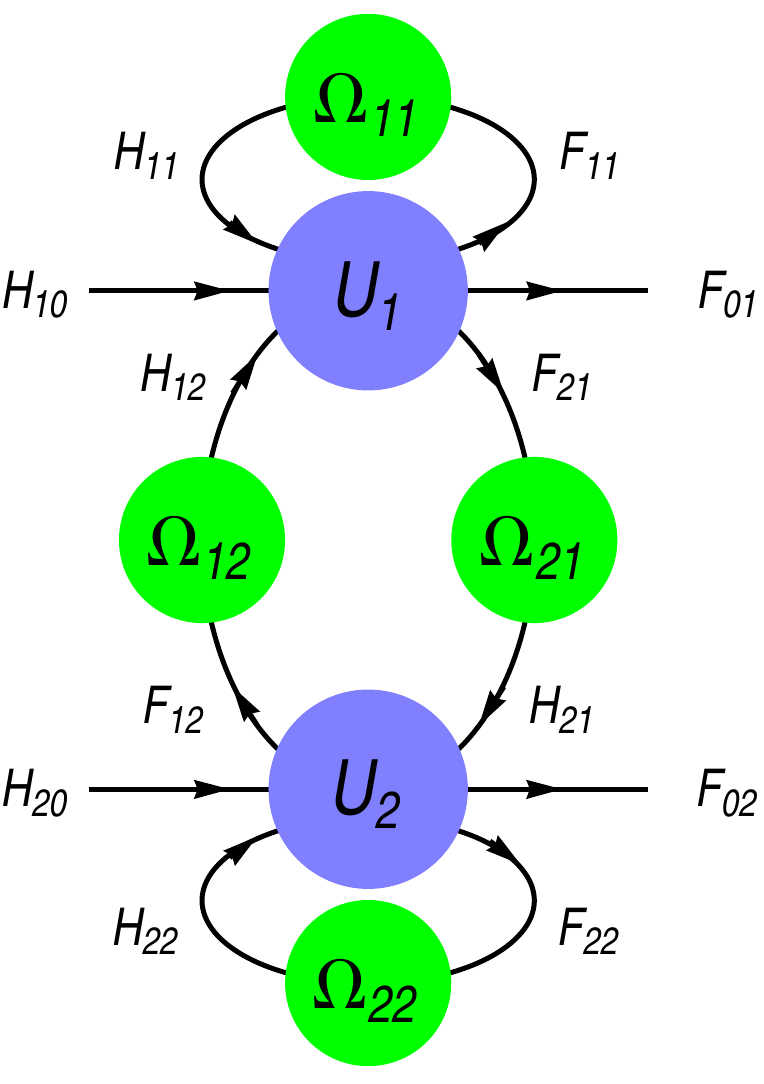}
    \end{center}
    \caption{A unitary network with two nodes. Some of the connections may be
      absent.}
    \label{fig:2}
\end{figure}

Several unitary operators can be reduced to form a network. This is
illustrated in Fig. \ref{fig:2} for a network with two nodes.

\newdefinition{df:2} ~{\em (Unitary network) } ~Let
$U_\alpha : \cH^{(\alpha)} \to \cF^{(\alpha)}$, with
$\alpha \in \{ 1, \ldots, N \}$ be a finite set of unitary operators. The
spaces admit the decompositions
\be
\cH^{(\alpha)} =
\Oplus_{\beta=0}^N \cH_{\alpha\beta},
\quad
\cF^{(\alpha)} =
\Oplus_{\beta=0}^N \cF_{\beta\alpha},
\quad
\alpha \in \{ 1, \ldots, N \}
\,.
\ee
Further
\be
\Omega_{\alpha\beta} :  \cF_{\alpha\beta} \to \cH_{\alpha\beta},
\qquad
\alpha,\beta \in \{ 1, \ldots, N \}
\ee
are unitary maps. The unitary network is defined as the reduction
$\OcU: \cH_0 \to \cF_0$, where
\be
U := \Oplus_{\alpha=1}^N U_\alpha
,\qquad
\Omega := \Oplus_{\alpha,\beta=1}^N \Omega_{\alpha\beta}
.
\ee
In this reduction
\bes
\cH &= \Oplus_{\alpha=1}^N \cH^{(\alpha)},
\qquad
\cF = \Oplus_{\alpha=1}^N \cF^{(\alpha)},
\\
\cH_0 &= \Oplus_{\alpha=1}^N \cH_{\alpha 0},
\qquad
\cF_0 = \Oplus_{\alpha=1}^N \cF_{0 \alpha},
\\
\cH_1 &= \Oplus_{\alpha,\beta=1}^N \cH_{\alpha\beta},
\qquad
\cF_1 = \Oplus_{\alpha,\beta=1}^N \cF_{\alpha\beta}.
\ees
\fin

The labels have been arranged with the convention that in $\alpha\beta$ the
flow goes from $\beta$ to $\alpha$.  Note that some (or even most) of the
subspaces may vanish\footnote{The unique map $0\to 0$ is unitary as it is a
  norm-preserving bijection.}, that is, the graph needs not be complete. Due
to unitarity of $U$'s and $\Omega$'s, the following conditions on the
dimension must be met:
\bes
&
\dim \cH_{\alpha\beta} = \dim \cF_{\alpha \beta}  \qquad \alpha,\beta >0
\,,
\\
&
\sum_{\beta \ge 0} \dim \cH_{\alpha \beta} =
\sum_{\beta \ge 0} \dim \cF_{\beta \alpha} 
\qquad \alpha >0
\,.
\ees

\section{ Application to quantum graphs }
\label{sec:3}

In this section we apply the reduction of unitary operators to the
scattering matrix of quantum graphs \cite{Kottos:1999,Kuchment:2008dub}. We
first consider a topological graph with a nonempty set $V'$ of vertices and a
nonempty set $E$ of edges, each edge joining two vertices. $|V'|$ and $|E|$
are finite and no vertices of degree (valence) $0$ are allowed. Momentarily it
will suffice to consider connected graphs as the extension to disconnected
graphs is straightforward.

The graph becomes a metric graph by assigning to each edge a positive length
that may be finite or infinite. The finite- and infinite-length edges will be
referred to as {\em internal} and {\em external lines},
respectively. Correspondingly $E = E_{\rm i} \cup E_{\rm e}$,
$E_{\rm i} \cap E_{\rm e} = \emptyset$. In the metric graph each line is a
mathematical line, a connected one-dimensional real smooth manifold, endowed
with a metric, and the points of the line belong to the metric graph.

There are two types of vertices, $V' = V \cup V_\infty$,
$V \cap V_\infty = \emptyset$.  The vertices in $V_\infty$ represent points at
infinity and have degree exactly one. All the points at infinity are regarded
as different. $V$ are the finite or regular vertices. The internal lines have
two vertices in $V$ as endpoints. For these an orientation is arbitrarily
adopted so that one of the vertices is the initial endpoint of the line and
the other is the final endpoint. Each point on the line $e\in E_{\rm i}$ has a
coordinate $x$ which the distance to the initial endpoint and $x\in [0,l_e]$
where $l_e$ is the finite length of the internal line.

External lines may have one or two endpoints in $V_\infty$. The latter
possibility would imply $|E_{\rm e}|=1$, $|V|=|E_{\rm i}|=0$ and will be disregarded.
Therefore we assume that all external lines are of semi-infinite type, with
one endpoint in $V$ and another in $V_\infty$ (hence $|V|>0$).  These lines
admit a canonical orientation, namely, from the regular vertex to the point at
infinity. The coordinate $x$ on the external line is the distance to the
regular-vertex endpoint and $x\in[0,+\infty)$.

Given a metric graph $\Gamma$, complex functions $f:\Gamma \to \C$ can be
defined, $p \mapsto f(p)$ where $p$ is any point on an internal or external
line, including the vertices. Here $\Gamma$ is regarded as a set including the
interior points in the lines and the regular vertices, but not the points at
infinity. The endpoints of different lines converging to a common vertex are
identified as a single point of the metric graph as a set. Then a Hilbert
space $L_2(\Gamma)$ can be defined.\footnote{Needless to say, this vector
  space is infinite-dimensional.} It is the orthogonal direct sum of all
$L_2(e)$, the measurable and square integrable functions on the edge $e$,
\be
\| f\|^2 := \sum_{e\in E} \int_0^{l_e} |f_e(x)|^2 dx \le +\infty
.
\ee

In its simpler version, the quantum graph is obtained by considering a non
relativistic quantum particle that moves on the metric graph. $\psi(p)$,
$p\in\Gamma$, denotes its wavefunction. This requires introducing a quantum
Hamiltonian $H$ as a suitable self-adjoint operator acting in $L_2(\Gamma)$.

We will only consider the potential-free case, hence on each line $e$ the
Hamiltonian operator is just $\ds -\frac{d^2}{dx^2}$ for interior points, with
domain the Sobolev space $H^2(e)$. Appropriate boundary conditions should be
given on the regular vertices to ensure self-adjointness. For each vertex
$v\in V$ the boundary conditions take the form
\be
A_v \Psi(v) + B_v \Psi'(v) = 0
.
\label{eq:2.2}
\ee
Here $\Psi(v)$ is the $d_v$-dimensional column vector of values of
$\psi_e(v)$, which are the wavefunctions at $p=v$ along the lines $e$ with
endpoints at $v$, and $d_v$ is the degree of the vertex $v$. The column vector
$\Psi'(v)$ contains $\psi'_e(v)$, the derivatives with respect to the distance
to the vertex. $A_v$ and $B_v$ are $d_v\times d_v$ constant complex
matrices. Then the Hamiltonian operator is self-adjoint if and only if i) the
$d_v\times 2 d_v$ matrix $(A_v,B_v)$ has rank $d_v$, and ii) the matrix
$A_v B_v^\dagger$ is Hermitian \cite{Berkolaiko:2013}.

Let $L:=|E_{\rm e}|=|V_\infty|$ be the number of external lines. For a compact
graph, i.e. $L=0$, the energy spectrum is discrete, however only the case $L>0$
will be considered as we are interested in the scattering problem.  Even if
there is no potential energy on the lines, the boundary conditions may
introduce contact interaction at the vertices, so the energy spectrum may have
a negative part. The scattering states have positive energy $k^2$ with $k>0$
and the spectrum will be degenerated in general. The eigenfunctions on each
line $e\in E$ are of the form $a e^{ikx} + b e^{-ikx}$, $a,b \in \C$, hence
there are $2|E|$ free coefficients. However each vertex $v\in V$ puts $d_v$
conditions. The graph identity $L+2|E_{\rm i}| = \sum_{v\in V} d_v$ then implies
that each eigenvalue $k^2$ has degeneracy $L$.

For $k>0$ let $\Psi(p)$ be the solution with wavefunctions
$a_e e^{ikx} + b_e e^{-ikx}$ for every $e \in E_{\rm e}$. The solution is
univocally determined by choosing $L$ values for the coefficients $b_e$, then
the linear map
\be
\forall e \in E_{\rm e} \qquad a_e = \sum_{e'\in E_{\rm e}} S(k)_{ee'} b_{e'}
.
\ee
defines the scattering matrix $S(k)$ of the quantum graph. This is a unitary
$L\times L$ matrix.

The simplest graphs are stars graphs, with just one vertex and $L$ external
lines. Since the wavefunctions on the lines are $a_e e^{ikx} + b_e e^{-ikx}$,
the boundary conditions in \nec{2.2}, can be cast in matrix form as
\be
A (a + b) + ik B(a-b) = 0
.
\ee
Solving $a = Sb$ leads to the scattering matrix for the star graph
\be
S = (ik B + A)^{-1}  (ik B - A)
.
\ee
The conditions stated previously on the matrices $A$ and $B$ guarantee that
$ik B + A$ is a regular matrix and $S$ is unitary \cite{Berkolaiko:2013}.

The discussion can be extended to disconnected graphs, including compact
components. The Hilbert space is the direct sum of the corresponding spaces
and the Hamiltonian is the sum of Hamiltonians of each subgraph. Likewise the
scattering matrix is the direct sum of the scattering matrices of the components
(counting the compact components as $0$-dimensional).

\newdefinition{df:3} ~{\em (Reduction of a quantum graph)} ~For a quantum
graph with $L\ge 2$, let $e_1$ and $e_2$ be two different external lines, with
corresponding regular vertices $v_1$ and $v_2$ (which need not be different).
A reduced graph is obtained by removing the two external lines (as well as
the corresponding points at infinity) and adding a new internal line $e$ with
vertices $v_1$ and $v_2$ and assigning a finite length $l_e$ to the new
internal line.  The reduced vertices may belong to different or the same
connected component.  The process decreases $L$ in two units. More generally,
a number $r$ of pairs of external lines may be removed to be replaced by
internal lines. In this case the number of external lines becomes $L-2r$ in
the reduced graph.  \fin

The reduction is fully specified by the choice of unordered pairs of
different external lines and the lengths of the resulting internal
lines. Consequently the same final graph is obtained when the reductions are
carried out sequentially in any order.

For a quantum graph $\Gamma$ and given $k>0$, let
$a_j e^{ikx} + b_j e^{-ikx}$, $\forall j\in E_{\rm e}$, be wavefunctions with
energy $k^2$ on the external lines. The coefficients $b_j$ and $a_j$ are the
amplitudes of the inward and outward waves, respectively. Let $\cH$ be the
Hilbert space isomorphic to $\C^L$ of the coefficients $b_j$, and similarly
$\cF$ the space of the coefficients $a_j$. The scattering matrix is then a
unitary map $S: \cH \to \cF$.

Let us consider a reduction of $\Gamma$, with pairings of external lines
$\{ (j_1,m_1), \ldots, (j_r,m_r) \}$ and lengths of the resulting internal
lines $\{ l_1,\ldots, l_r \}$, where $2r \le L$ is assumed. Then
$\cH = \cH_0 \oplus \cH_1$ and $\cF = \cF_0 \oplus \cF_1$, where $\cH_1$
contains the coefficients $b_{j_\ell}$ and $b_{m_\ell}$,
$\ell\in\{1,\ldots,r\}$ and $\cF_1$ contains the coefficients $a_{j_\ell}$ and
$a_{m_\ell}$, i.e. those corresponding to the external lines to be
reduced. $\dim\cH_1 = \dim \cF_1 = 2r$. In turn $\cH_0$ and $\cF_0$ are
spanned by the inward and outward coefficients of the unreduced external
lines. A $2r\times 2r$-dimensional matrix $\Omega$ can be constructed as a
direct sum of $r$ \,$2\times 2$ blocks, as follows
\be
\Omega  = \Oplus_{\ell=1}^r\PM{
0 & \omega_\ell \\ \omega_\ell & 0
},
\qquad
\omega_\ell := e^{ik l_\ell}
\quad  
\ell \in \{1,\ldots, r \}
.
\ee
$\Omega$ is a unitary operator from $\cF_1$ to $\cH_1$. The blocks connect the
two spaces using the scheme
\be
\PM{ b_{j_\ell} \\ b_{m_\ell} }
=
\PM{
0 & \omega_\ell \\ \omega_\ell & 0
}
\PM{ a_{j_\ell} \\ a_{m_\ell} }
\qquad  
\ell \in \{1,\ldots, r \}
.
\ee

\newtheorem{th:5} ~Let $\Gamma$ be a quantum graph with scattering matrix $S$,
and let $\Gamma^{\,\prime}$ be the reduced graph with connections as
specified in $\Omega$, with scattering matrix $S'$. Then
$S' = \Omega \cdot S$.

\proof ~Because the reduction of graphs and the reduction of unitary
operators can be carried out sequentially it will be sufficient to consider the
case $r=1$, i.e., just two different external lines $j,m\in E_{\rm e}$ are
reduced, to yield an internal line of length $l$.

Let $S$ be the scattering matrix of the original graph $\Gamma$. (More
precisely the original graph removing its compact components, which play no
role for the scattering but introduce a discrete spectrum at positive
energies.)  Starting from the $2|E|$ free coefficients $a_e,b_e$, $e\in E$,
the boundary conditions at the vertices impose $\sum_{v\in V} d_v$
constraints, leaving $L$ unconstrained parameters, which can be chosen as the
$b_e$, $e\in E_{\rm e}$. In the reduced graph $\Gamma^{\,\prime}$ the lines $j$ and
$m$ are identified, this implies the identification of the corresponding
wavefunctions:
\be
a_j e^{ikx} + b_j e^{-ikx} =
a_m e^{ik(l-x)} + b_m e^{-ik(l-x)}
\ee
where $x$ is the distance to the regular vertex of $j$. Therefore, the
operator $S'$ (the scattering matrix of $\Gamma^{\,\prime}$) is defined by the same
equations as $S$ by adding the new constraints
\be
b_j = a_m \omega
,\qquad
b_m = a_j \omega
, \qquad \omega := e^{ikl}
.
\ee
All the coefficients corresponding to internal lines of $\Gamma$ are fully
fixed by the equations, therefore we can retain explicitly just those of the
external lines.  Let us denote $\psi \in \cH$ the column vector of the $L$
coefficients $b_e$, $e \in E_{\rm e}$, and similarly $\phi \in \cF$ for the
$a_e$. These vectors can be split as $\psi_0+\psi_1$ and $\phi_0+\phi_1$,
where $\cH_1$ refers to $b_j$ and $b_m$ and $\cF_1$ to $a_j$ and $a_m$. Then
the equations defining $S'$ are
\be
\phi_0+\phi_1 = S( \psi_0+\psi_1 ),
\qquad
\psi_1 = \Omega \phi_1
,
\ee
where $\ds \Omega = \PM { 0 & \omega \\ \omega & 0}$. 
In $\Gamma^{\,\prime}$ the scattering matrix is the map $\psi_0\to\phi_0$,
thus $S' = \Omega \cdot S$. \qed

\begin{figure}[ht]
  \begin{center}
    \includegraphics[height=40mm]{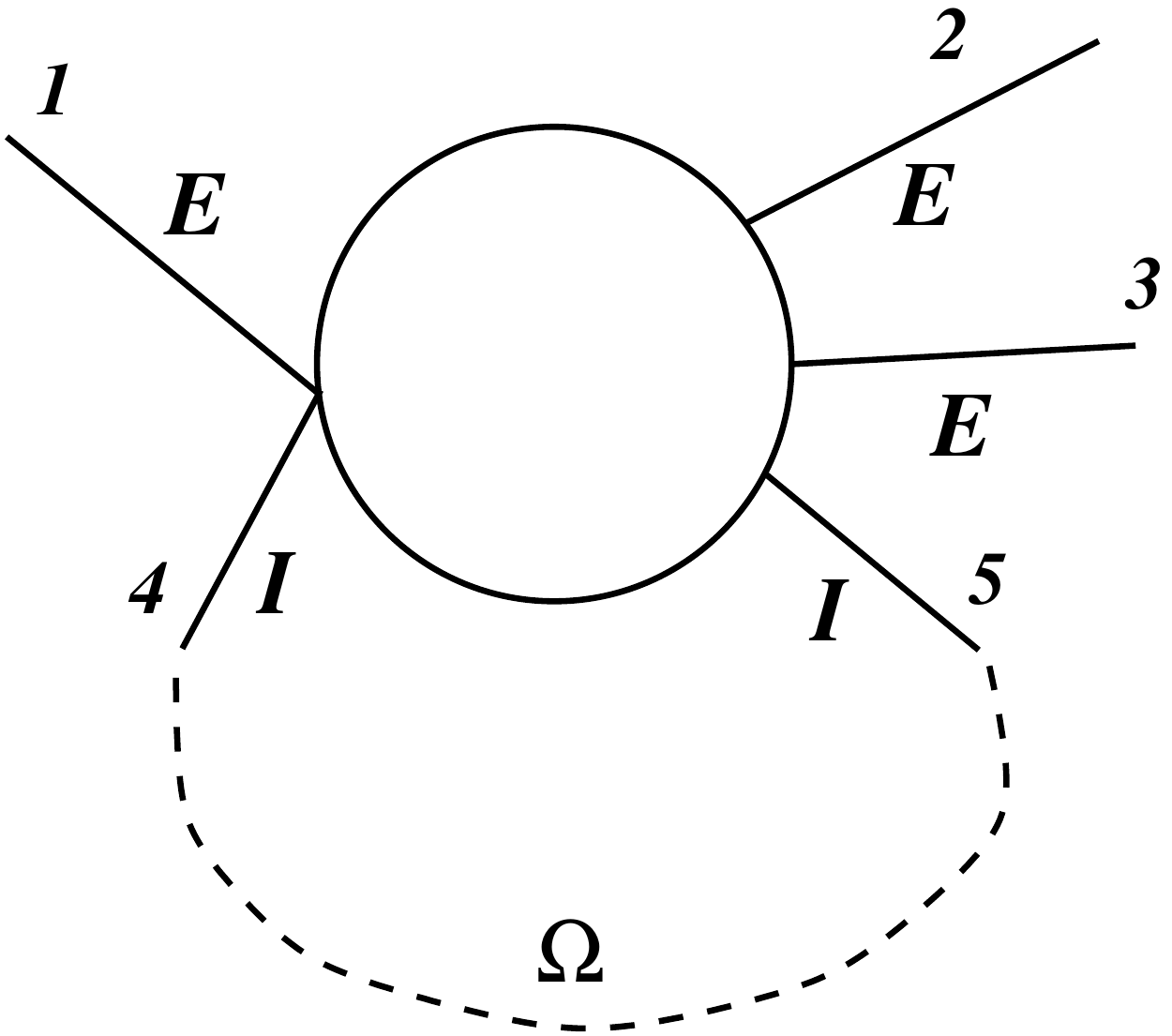}
    \end{center}
\caption{Reduction of a graph. The external lines $4$ and $5$ are
reduced and the new line is assigned a length $ l$, so that the phase is
$\omega = e^{ikl}$. The lines $1$, $2$ and $3$ remain external so they carry a
label $E$, while $4$ and $5$ become internal and carry a label $I$. }
\label{fig:3}
\end{figure}

An example of reduction is displayed in Fig. \ref{fig:3}. Initially the
scattering matrix has dimension $5$. The space can be split into an external
sector, with label $E$, corresponding to lines $1$, $2$ and $3$, and a
internal sector, with label $I$, corresponding to lines $4$ and $5$. The
scattering matrix then gets decomposed into submatrices as
\be
S = \PM{ S_{EE} & S_{EI} \\ S_{IE} & S_{II}   }
.
\ee
In the present example the blocks $S_{EE}$, $S_{II}$, $S_{EI}$ and $S_{IE}$,
have dimensions $3\times 3$, $2\times 2$, $3\times 2$ and $2\times 3$,
respectively, while the matrix $\Omega$ is $2\times 2$. From Theorem
\ref{th:5}, and making use of \nec{1.10a}, it follows that the scattering
matrix of the reduced graph can be expressed as
\be
S' = S_{EE} + S_{EI} (\Omega^{-1} - S_{II} )^{-1} S_{IE}
.
\label{eq:2.12}
\ee
Of course this formula holds in general, not just in our example.
A similar expression has been derived previously in \cite{Caudrelier:2009ay}.

As we have noted above, every vertex defines a unitary scattering matrix
(regarding the vertex as a star graph) and obviously every graph can be
obtained through reduction of the star graphs defined by its
vertices. Therefore, \nec{2.12} provides the scattering matrix for an
arbitrary graph.

\section{ Non unitary reductions }
\label{sec:2}

\subsection{ Reduction of non-unitary operators }
\label{sec:2.a}
  
Under suitable conditions the notion of reduction of an operator with a
suboperator can be extended to more general operators and spaces. Let
$\cH = \cH_0 \oplus \cH_1$ and $\cF = \cF_0 \oplus \cF_1$ be complex and
finite dimensional vector spaces (the direct sum no longer implies
orthogonality) and let $A: \cH \to \cF$ and $B:\cF_1 \to \cH_1$ be linear
maps. Relations of the type $\dim \cH = \dim \cF$, etc, no longer need to
hold. One can still pose the problem whether the equations
\bes
&
\phi_0 + \phi_1 = A (\psi_0 + \psi_1), \qquad \psi_1 = B \phi_1,
\\
&
\psi_j \in \cH_j, \quad
\phi_j \in \cF_j, \quad j\in\{0,1\}
\label{eq:1.38}
\ees
with input $\psi_0$ and unknowns $\psi_1$, $\phi_0$ and $\phi_1$, define a
linear map $\psi_0 \to \phi_0$. At variance with the unitary case, for more
general operators this is not always guaranteed.

\newproposition{pr:6a} ~The following are necessary and sufficient conditions
for the \eqs{1.38} to define a linear map $\psi_0 \to \phi_0$:
\begin{itemize}
\item[i)] $F_1 A \cH_0 \subseteq (I - F_1 A B ) \cF_1$ \quad(existence)
\item[ii)] $F_0 AB \cV_1 = 0$ \quad(uniqueness of $\phi_0$)
\end{itemize}
$H_j$ and $F_j$, for $j\in\{0,1\}$, denote the projectors onto $\cH_j$ and
$\cF_j$, respectively, and
\be
\cV_1 := \{ \varphi_1 \in \cF_1 ~|~ F_1 A B \varphi_1 = \varphi_1 \}
.
\label{eq:3.3}
\ee

\proof ~Upon projection of \nec{1.38} onto $\cF_0$ and $\cF_1$, and elimination
of $\psi_1$, the equations become
\be
\phi_0  = F_ 0 A (\psi_0 +  B \phi_1),
\qquad
\phi_1  = F_ 1 A (\psi_0 +  B \phi_1)
.
\ee
From the second equation it follows that the condition i) is necessary and
sufficient for the existence of $\phi_1$ for all $\psi_0$, and then of
$\phi_0$, using the first equation. $\cV_1$ is the kernel of the operator
$(I - F_1 AB) : \cF_1 \to \cF_1$. A component $\varphi_1 \in \cV_1$ introduces
an additive ambiguity in $\phi_1$. When inserted in the first equation,
condition ii) encodes that $\phi_0$ is independent of $\varphi_1$, and so
unique. \qed

Obviously both conditions are met when $\cV_1 = 0$, so this is a sufficient
condition for \eqs{1.38} to define a map $\psi_0\to\phi_0$. While sufficient,
this condition is not necessary, the unitary case providing a
counterexample.\footnote{In the unitary case the spaces $\cV_1$ defined in
  \nec{2.5} and \nec{3.3} coincide.}

The requirement of existence and uniqueness of the map suggests a possible
definition of the reduction $B \cdot A$, which would fulfill the relation
\be
B\cdot A = F_0 (I - A B F_1 )^{-1} A H_0
\,.
\label{eq:1.43}
\ee
But this is not the only possibility. The reduction in the unitary case was
also achieved as the sum of a series. An iterative treatment of the \eqs{1.38}
would take the form
\bes
A \psi_0 &=  \phi_0^{(1)} + \phi_0^{(1)},
\\
A B  \phi_1^{(n)} &= \phi_0^{(n+1)} + \phi_1^{(n+1)} \quad \forall n \ge 1,
\\
\phi_0 &= \sum_{n=1}^\infty \phi_0^{(n)},
\qquad
\phi_1 = \sum_{n=1}^\infty  \phi_1^{(n)}
.
\label{eq:3.6a}
\ees
Clearly when both series are absolutely convergent they produce a solution of
the equations, and hence a well-defined map $\psi_0 \to \phi_0$. By absolute
convergence it is meant convergence of the series
\be
\sum_{n=1}^\infty  \| \phi_1^{(n)} \|
\ee
(and similarly for $\phi_0$) using any vector-norm in $\cF$. The definition
does not depend on the concrete norm chosen. This follows from the property of
finite-dimensional vector spaces that for any other norm $\| ~ \|'$, there
exist a $K>0$ such that $\forall v \in \cF ~ \| v \|' \le K \| v \|$. Another
observation is that absolute convergence of $\phi_1$ automatically implies
that of $\phi_0$, due to
\be
\| \phi_0^{(n+1)} \| = \| F_0 A B  \phi_1^{(n)} \| \le \| F_0 A B \| \|
\phi_1^{(n)} \|
.
\ee

The two possible definitions of $B \cdot A$ (namely, by uniqueness of the
solution of the equation with respect to $\phi_0$, and by absolute convergence
of the series) are different in general. There are pairs $(A,B)$ such that the
solution is unique for $\phi_0$ with a divergent series, and it is also
possible to violate the condition ii) in Proposition \ref{pr:6a} having an
absolutely convergent series. We will adopt the definition of $B \cdot A$
based in series as the more useful one, giving up the requirement of
uniqueness of the solution for $\phi_0$.

\newdefinition{df:7} ~{\em (Reduction of an operator with a suboperator)}
~The reduction $B \cdot A : \cH_0 \to \cF_0$ is defined as the linear map
$\psi_0\to \phi_0$ determined by \eqs{3.6a} whenever the series defining
$\phi_1$ is absolutely convergent. Such $\phi_0$ and $\phi_1$ solve
\eqs{1.38}. The existence of additional solutions of the equations with a
different $\phi_0$ is not excluded.  \fin

The series defining the reduction is compactly expressed in \eq{1.43}.

\newproposition{pr:13} ~When it is well-defined, the reduction fulfills
\be
B \cdot A = \lim_{ n \to \infty } (F_0 + A B F_1 )^n A H_0
\,.
\label{eq:1.43a}
\ee

\proof ~The relations $F_j F_k = \delta_{jk} F_j$ ~($j,k \in \{0,1\}$) ~yield the
identity
\be
(F_0 + A B  F_1 )^n A H_0 =
F_1 (A B F_1)^n A H_0 +  \sum_{k=0}^n F_0 (A B F_1)^k A H_0
\,.
\ee
When $n\to\infty$, the first term in the RHS goes to $0$ due to the absolute
convergence of the series for $\phi_1$ in \eq{3.6a}. The second term
reproduces the series defining $\phi_0$. \qed

\newtheorem{th:6} ~A sufficient condition for the reduction $ B \cdot A$ to
be well-defined is the existence of a subspace $\cW \subseteq \cF_1$ with the
following properties:
\begin{itemize}
\item[i)] $F_1 A \cH_0 \subseteq \cW$,
\item[ii)] $ F_1 AB \cW \subseteq \cW$,
\item[iii)] $ \| (F_1 AB)^k \|_{\cW} < 1$ for some $k\ge 1$, where
  $\|~\|_{\cW}$ denotes the norm restricted to $\cW$, and $\|~\|$ denotes the
  operator-norm induced by some vector-norm in $\cF$.
\end{itemize}

\proof ~Under the condition i) $\phi_1^{(1)} \in \cW$ and ii) implies that
$\phi_1^{(n)} \in \cW$ $\forall n \ge 1$. Then iii) guarantees the geometric
convergence of the series: for a general term $n = m k+r$, with $m,r\in\Z$ and
$r \in \{0,\ldots,k-1\}$, then
$\| (F_1 AB)^n \|_{\cW} \le \| (F_1 AB)^k \|_{\cW}^m \| F_1 AB \|_{\cW}^r$, so
each subseries $r=0,\ldots, k-1$ is absolutely convergent and $B \cdot A$ is
well-defined.  \qed

When $\cW = \cF_1$ necessarily $\cV_1 = 0$, and uniqueness of the solution of
\eqs{1.38} with respect to $\phi_0$ is also guaranteed.

\subsection{ Reduction of quantum channels }

\newdefinition{df:4} ~{\em (Quantum channel)} ~Let $\cH$ and $\cF$ be Hilbert
spaces. Let $\cL(\cH) = \cH \otimes \cH^*$ and $\cL(\cF) = \cF \otimes \cF^*$
denote the linear operators (endomorphisms) in $\cH$ and in $\cF$,
respectively. $T$ is a {\em quantum channel} from $\cH$ to $\cF$ when it is a
linear map $T: \cL(\cH) \to \cL(\cF)$ (a superoperator) that is completely
positive (CP) and trace preserving (TP). \fin

In order to extend the concept of reduction to quantum channels, let
$\cH = \cH_0 \oplus \cH_1$ and $\cF = \cF_0 \oplus \cF_1$, and let
$T: \cL(\cH) \to \cL(\cF)$ and $R: \cL(\cF_1) \to \cL(\cH_1)$ be quantum
channels. $T$ is the CPTP superoperator to be reduced and $R$ the
connection and we would like to give a meaning to the reduction $R\cdot T$,
as a quantum channel from $\cH_0$ to $\cF_0$.

A natural desideratum would be to demand that when $T$ and $R$ are unitary
channels, corresponding to unitary maps $U$ and $\Omega$, $R\cdot T$ would be
a unitary quantum channel for $\Omega\cdot U$, however, this requirement
cannot be met. To be more specific, let us introduce the notation
\cite{Cai:2019}
\be
\supop{A}(~) := A (~) A^\dagger,
\ee
where $A$ is an operator and so $\supop{A}$ is a superoperator.  The symbol
fulfills the property $\supop{AB} = \supop{A}\,\supop{B}$.  By definition, $T$
is a unitary quantum channel for the unitary operator $U$ when
$T = \supop{U}$, that is, $T(\rho) = U \rho U^\dagger$. Likewise
$R = \supop{\Omega}$. Then the requirement (where the LHS is yet to be
defined)
\be
\supop{ \Omega} \cdot \supop{U} = \supop{ \Omega \cdot U} 
\ee
is inconsistent, because $\supop{U}$ and $\supop{\Omega}$ 
are invariant under $U \to \omega U$, $\Omega \to \omega' \Omega$
($\omega,\omega'\in \C$ and $|\omega|=|\omega'|=1$) while
$\supop{ \Omega \cdot U}$ is not, unless $\omega'=\omega^*$. (In general
$(\omega' \Omega ) \cdot (\omega U) $ is not $\omega'' (\Omega \cdot U)$.)

A possible definition for $R\cdot T$ comes from Stinespring's dilation theorem
\cite{Paulsen:2003}, which associates every CPTP superoperator to a unitary
operator in an extended Hilbert space. Unfortunately the unitary operators $U$
and $\Omega$ corresponding to $T$ and $R$ are not unique and the different
choices would yield different results for $\Omega \cdot U$. Therefore the
definition of $R\cdot T$ would not depend solely on the channel $T$ and the
subchannel $R$. So we disregard this approach.

Let us decompose $\cL(\cH)$ into the four sectors,
\be
\cL(\cH) = \cH_0 \otimes \cH_0^* \oplus
\cH_0 \otimes \cH_1^* \oplus
\cH_1 \otimes \cH_0^* \oplus
 \cH_1 \otimes \cH_1^* 
 ,
\ee
and similarly for $\cL(\cF)$. Then, the operators $\rho \in \cL(\cH)$ and
$\sigma \in \cL(\cF)$ admit the decompositions
\bes
\rho &= \rho_{00} + \rho_{01} + \rho_{10} + \rho_{11},
\\
\sigma &= \sigma_{00} + \sigma_{01} + \sigma_{10} + \sigma_{11}
.
\ees
In analogy with \nec{1.38}, we can pose the problem whether, for given $T$ and
$R$, and $\rho_{00}$, the equations
\be
\sigma = T(\rho),
\qquad
R(\sigma_{11}) = \rho_{11}
,
\qquad
\rho_{01}= \rho_{10} = 0
,
\label{eq:2.12a}
\ee
have a solution for $\rho_{01}$, $\rho_{10}$, $\rho_{11}$, $\sigma_{00}$,
$\sigma_{01}$, $\sigma_{10}$ and $\sigma_{11}$, unique for $\sigma_{00}$,
hence providing a well-defined linear map $\rho_{00}\to \sigma_{00}$.

Note that choices have been made in writing the equations in \nec{2.12a},
namely, the conditions $\rho_{01}= \rho_{10} = 0$.  This is necessary (or at
least convenient) to match the number of unknowns with that of equations. The
choice taken corresponds to the iterative process to be set up below, in
\nec{2.22}. The equations \nec{2.12a} do not distinguish between a quantum
channel $T$ and its ``decohered'' version $T'$
\be
T' := (\supop{F_0} + \supop{F_1}) \mcirc T \mcirc (\supop{H_0} + \supop{H_1})
.
\ee
(As before, $H_j$ and $F_j$, with $j\in\{0,1\}$, denote orthogonal
projectors.)  $T'$ is also a quantum channel which only attends to the
diagonal sectors of $T$. That is,
\be
T'(\rho_{01} + \rho_{10} ) = 0
\ee
and
\be
\sigma_{00} + \sigma_{11}
= T'(\rho_{00} + \rho_{11})
\ee
when
\be
\sigma_{00} + \sigma_{01} + \sigma_{10} + \sigma_{11}
=
T(\rho_{00} + \rho_{11}),
\ee
Therefore any information related to coherence between the spaces $\cH_0$ and
$\cH_1$, or $\cF_0$ and $\cF_1$ is lost in the map $\rho_{00}\to\sigma_{00}$
defined by \nec{2.12a}. It follows that the set of solutions for the pairs
$(T,R)$ or $(T',R)$ are identical.

Since the off-diagonal spaces $\cH_0\otimes\cH_1^*$, $\cH_1\otimes\cH_0^*$,
$\cF_0\otimes\cF_1^*$ and $\cF_1\otimes\cF_0^*$ do not play a role, in what
follows we will use the notation $\rho_0$ (for $\rho_{00}$) and $\sigma_0$ and
$\sigma_1$ (for $\sigma_{00}$ and $\sigma_{11}$ respectively). In this way
\nec{2.12a} can be expressed as
\bes
&
\sigma_0 + \sigma_1 = T'(\rho_0 + \rho_1),
\qquad
R (\sigma_1) = \rho_1
,
\\
&
\rho_j \in \cL(\cH_j),
\quad
\sigma_j \in \cL(\cF_j) \quad j \in \{0,1\}
.
\label{eq:2.12c}
\ees

It must be noted that, unlike what happened in the reduction of unitary
operators, that $T$ and $R$ are quantum channels does not guarantee that
\eqs{2.12c} have a solution. A counterexample is readily found: let
$\cH = \cF = \C^2$, with $ \cH_0=\cF_0 = \{ (z,0) ~|~ z\in\C \}$, $R$ the
identity map and
\be
T(\rho) := \Tr(\rho) H_1 \,.
\label{eq:2.15}
\ee
As can be verified, for such $T$ and $R$, \eqs{2.12c} are inconsistent unless
$\rho_0 = 0$.

Even though \eqs{2.12c} are formally identical to \eqs{1.38} (viewing the
quantum channels $T'$ and $R$ as (super)operators, and the operators $\rho$
and $\sigma$ as (super)vectors) the treatment cannot be identical due to the
additional requirement that the linear map $\rho_0 \to \sigma_0$ must be CP
(as well as TP, but this is a straightforward consequence of the
equations). Positivity is a crucial issue in the context of quantum channels,
and the treatment has to be adapted accordingly.

For convenience, out of $T$ and $R$, we define the following CP
superoperators,
\bes
T_{j0} &:= \supop{F_j}\mcirc T \mcirc \supop{H_0} : \cL(\cH_0) \to \cL(\cF_j),
\qquad
\\
\tT_{j1} &:= \supop{F_j} \mcirc T \mcirc R :\cL(\cF_1) \to \cL(\cF_j),
\qquad j \in \{0,1\}
.
\label{eq:2.58}
\ees
Note that $T_{00} + T_{10}$ and $\tT_{01} + \tT_{11}$ are both CPTP, and also
that the same operators are produced using $T'$ instead of $T$. An iterative
treatment, aiming to a series solution of \eqs{2.12c}, produces
\bes
\sigma_0^{(1)} &= T_{00}(\rho_0),
\qquad
\sigma_1^{(1)} = T_{10}(\rho_0),
\\
\sigma_0^{(n+1)} &= \tT_{01}(\sigma_1^{(n)}),
\qquad
\sigma_1^{(n+1)} = \tT_{11}(\sigma_1^{(n)})
\quad \forall n \ge 1,
\\
\sigma_0 &= \sum_{n=1}^\infty \sigma_0^{(n)}
,
\qquad
\sigma_1 = \sum_{n=1}^\infty \sigma_1^{(n)}
.
\label{eq:2.22}
\ees
If $\rho_0$ is a positive operator (more precisely, a non-negative operator,
$\rho_0 \ge 0$) the operators $\sigma_0^{(n)}$ and $\sigma_1^{(n)}$ are all
positive too. If both series in \nec{2.22} are absolutely convergent,
$\sigma_0$ and $\sigma_1$ provide a solution to \nec{2.12c} and a well-defined
positive linear map $\rho_0 \to \sigma_0$.

In order to give a precise meaning to the convergence of the series, the
operator-norm induced in $\cL(\cF)$ by the Hilbert space-norm is not a
natural one for quantum channels. Instead the trace is more
appropriate. To this end we introduce the following definition:

\newdefinition{df:6} ~For positive operators $\rho \in \cL(\cH)$ (actually
non-negative, $\rho\ge 0$) let $\|\rho\|_t := \Tr(\rho)$. The function
$\|~\|_t$ enjoys the required properties of a norm: $\|\rho\|_t\ge 0$ and
$\|\rho\|_t = 0$ only if $\rho=0$, and
$\| \lambda \rho\|_t = \lambda \| \rho \|_t$ for $\lambda \ge 0$, as well as
$\| \rho + \sigma \|_t= \| \rho \|_t + \| \sigma \|_t$. In turn, this induces
a norm on positive superoperators. For a linear $T:\cL(\cH)\to \cL(\cF)$ and
$T \ge 0$,
\be
\| T \|_t := \sup \{ \Tr(T(\rho)) ~|~ \forall
\rho\in\cL(\cH), ~\rho\ge 0, ~\Tr(\rho)=1 \}
.
\ee
This induced norm has the basic properties
$\| T_1 \mcirc T_2 \|_t \le \| T_1 \|_t \| T_2 \|_t$ and
$\|T_1 + T_2 \|_t \le \| T_1 \|_t+ \| T_2 \|_t$.  \fin

With this definition $\| T\|_t = 1$ when $T$ is a quantum channel. 

\newlemma{lm:4} ~Let $E^{(i)}: \cL(\cH) \to \cL(\cF)$, $i\in \{1,\ldots,n\}$,
be completely positive superoperators, and $E= \sum_{i=1}^n E^{(i)}$ a quantum
channel. Let $S^{(i)}: \cL(\cF) \to \cL(\cG)$, $i\in \{1,\ldots,n\}$, be
quantum channels. Then
\be
T = \sum_{i=1}^n S^{(i)} \mcirc E^{(i)}
\ee
is a quantum channel.

\proof ~The composition and sum of completely positive operators is completely
positive. Regarding the preservation of the trace
\bes
\Tr_{\cG}(T(\rho)) &=
\sum_{i=1}^n \Tr_{\cG} (S^{(i)} (  E^{(i)} (\rho) ) )
=
\sum_{i=1}^n \Tr_{\cF}  (  E^{(i)} (\rho) )
\\ &=
\Tr_{\cF} (E(\rho)) = \Tr_\cH(\rho)
.
\ees
\qed

\newproposition{pr:14} ~If the series of $\sigma_1$ in \eqs{2.22} is
absolutely convergent (in the sense of the norm $\|~\|_t$):
\begin{itemize}
\item[i)] The series of $\sigma_0$ is also  absolutely convergent.
  \item[ii)] The map $\rho_0 \to \sigma_0$ is CPTP.
  \end{itemize}
  
\proof ~The assertion i) follows from
\be
\| \sigma_0^{(n+1)} \|_t = \| \tT_{01}(\sigma_1^{(n)}) \|_t \le
\| \tT_{01} \|_t \| \sigma_1^{(n)} \|_t
.
\ee
To prove ii), let us establish the relation (provisionally denoting
$R \cdot T$ the map $\rho_0 \to \sigma_0$)
\be
R \cdot T =
\lim_{n\to\infty} \tT^n \mcirc T' \mcirc \supop{H_0},
\qquad
\tT := \supop{F_0} + T' \mcirc R \mcirc  \supop{F_1}
\,.
\label{eq:2.18}
\ee
The repeated application of $\tT$ produces
\be
\rho_0 \to \tT^n \mcirc T' (\rho_0) =
\sum_{k=1}^n \sigma_0^{(k)} + \sigma_1^{(n)} 
\quad
\forall n \ge 1 .
\ee
By Lemma \ref{lm:4}, $\tT$ is a quantum channel, hence for finite $n$,
$\tT^n \mcirc T' \mcirc \supop{H_0}$ is a quantum channel from $\cH_0$ to
$\cF$. The limit exists due to the absolute convergence, and yields the map
$\rho_0 \to \sigma_0$ which is then a quantum channel from $\cH_0$ to $\cF_0$.
\qed

We can now give a proper definition for the reduction $R \cdot T$,

\newdefinition{df:5} ~{\em (Reduction of quantum channels)} ~Given $T$ and
$R$, and $\rho_0$, the reduction $R \cdot T$ is defined as the map
$\rho_0 \to \sigma_0$ determined by \eqs{2.22}, provided the series for
$\sigma_1$ is absolutely convergent in the sense of $\|~\|_t$.
By Proposition \ref{pr:14} this property guarantees that $\sigma_0$ is also
absolutely convergent and the map $\rho_0 \to \sigma_0$ is a quantum channel.

Clearly with this definition the reduction depends only on $T$ and $R$
themselves, and $ R \cdot T = R \cdot T'$. Also, the $\rho$ and $\sigma$
defined by the series solve \eqs{2.12c}.  \fin

The series defining the reduction can be
expressed as
\be
R \cdot T =
 T_{00} + \tT_{01} \mcirc ( I + \tT_{11} + \tT_{11}^2+ \tT_{11}^3+ \cdots )
 \mcirc T_{10}
,
\label{eq:2.23b}
\ee
or more compactly as
\bes
R \cdot T &= T_{00} + \tT_{01} \mcirc ( I - \tT_{11})^{-1} \mcirc T_{10}
\\ &=
\supop{F_0} \mcirc (I - T \mcirc R \mcirc \supop{F_1} )^{-1}
\mcirc T \mcirc \supop{H_0}
,
\label{eq:2.23a}
\ees

\newtheorem{th:7} ~A sufficient condition for the reduction $ R \cdot T$ to
be well-defined is the existence of a subspace
$\hat{\cW} \subseteq \cL(\cF_1)$ with the following properties:
\begin{itemize}
\item[i)] $T_{10} \cL(\cH_0) \subseteq \hat{\cW}$,
\item[ii)] $ \tT_{11} \hat{\cW} \subseteq \hat{\cW}$,
\item[iii)] $ \| (\tT_{11})^k \|_{t,\hat{\cW}} < 1$ for some $k\ge 1$, where
  $\|~\|_{t,\hat{\cW}}$ denotes the norm $\|~\|_t$ restricted to $\hat{\cW}$.
\end{itemize}

\proof ~Conditions i) and ii) guarantee that $\sigma_1^{(n)} \in
\hat{\cW}$. Condition iii) guarantees a geometric convergence rate of the
series. \qed

\newexample{ex:2} ~Let us consider a simple example.  Let
$U_{jk}^{(m)}:\cH_k \to \cF_j$, $\forall m\in I_{jk}$ (a finite index set)
for $j,k \in \{0,1\}$, be unitary operators. And also unitary
$\Omega^{(n)}:\cF_1\to \cH_1$, $\forall n \in I'$ (another finite index set).
In addition, let
\bes
&
p_{jk}^{(m)}\ge 0, \qquad q^{(n)}\ge 0,
\qquad
\bar{p}_{jk}:= 
\sum_m p_{jk}^{(m)},
\qquad
\\
&
\bar{p}_{00} + \bar{p}_{10} = 1,
\qquad
\bar{p}_{01} + \bar{p}_{11} = 1,
\qquad
\sum_n q^{(n)} = 1
\ees
Then
\be
T_{jk} := \sum_m p_{jk}^{(m)} \supop{ U_{jk}^{(m)} },
\qquad
T := \sum_{j,k} T_{jk},
\qquad
R := \sum_n q^{(n)} \supop{ \Omega^{(n)} }
.
\ee

$R$ is a random unitary channel, i.e., a convex combination of unitary
channels.  $T$ is closely related to, but not exactly, a random unitary
channel, as it is not a mixture of unitary operators $\cH\to \cF$ (except for
special choices of the weights $p_{jk}^{(m)}$).  $T$ has been chosen in such
a way that $T=T'$.

\Eq{2.12c} corresponds to
\bes
\sigma_0 &= \sum_m p_{00}^{(m)} \supop{ U_{00}^{(m)} } \rho_0
+
\sum_m p_{01}^{(m)} \supop{ U_{01}^{(m)} } \sum_n q^{(n)} \supop{ \Omega^{(n)}
} \sigma_1
,
\\
\sigma_1 &= \sum_m p_{10}^{(m)} \supop{ U_{10}^{(m)} } \rho_0
+
\sum_m p_{11}^{(m)} \supop{ U_{11}^{(m)} } \sum_n q^{(n)} \supop{ \Omega^{(n)}
} \sigma_1
.
\label{eq:3.30}
\ees
Attending to the traced equations
\bes
\Tr(\sigma_0) &= \bar{p}_{00} \Tr(\rho_0)
+
\bar{p}_{01} \Tr( \sigma_1 )
,
\\
\Tr(\sigma_1) &= \bar{p}_{10} \Tr(\rho_0)
+
\bar{p}_{11} \Tr( \sigma_1 )
,
\ees
two cases can be distinguished:
\begin{itemize}

\item[i)] $\bar{p}_{11} = 1$. Since we want a generic $\rho_0$, the system can
  only be consistent if $\bar{p}_{10} = 0$. Then also $\bar{p}_{00} = 1$, and
  $p_{01}^{(m)} = p_{10}^{(m)} = 0$ $\forall m$.  The two sectors $0$ and $1$ are
  decoupled. $T_{10}=0$ and the series converges in one step.  $R\cdot T$ is
  just $T_{00}$.

\item[ii)] $\bar{p}_{11} < 1$. In this case
  $\ds \Tr(\sigma_1) = \frac{\bar{p}_{10}}{ \bar{p}_{01}}
  \Tr(\rho_0)$. $\sigma_1$ is not bounded for fixed $\Tr(\rho_0)$, but still
  $\Tr(\sigma_0) = \Tr(\rho_0)$ as it should. For the solution to the full
  (untraced) equations \nec{3.30} the condition $\bar{p}_{11} < 1$ ensures the
  convergence of the series for $\sigma_1$, or equivalently of that in
  \nec{2.23b}. The latter expression shows that, in this example, $R \cdot T$
  is a random unitary quantum channel (with an infinite, but convergent, Kraus
  representation).
\end{itemize}
\fin

The reduction of unitary quantum channels always produces a quantum channel:

\newproposition{pr:12} ~Let $U: \cH_0 \oplus \cH_1 \to \cF_0 \oplus \cF_1$ and
$\Omega: \cF_1 \to \cH_1$ be unitary operators. Then the reduction of the
unitary quantum channels $\supop{\Omega}$ and $\supop{U}$ exists and defines a
quantum channel. Moreover,
\be
\supop{\Omega} \cdot \supop{U}  = T_{U,\Omega}
,
\ee
where  $T_{U,\Omega}$ is the quantum channel of \eq{2.59}.

\proof ~For $T = \supop{U}$ and $R = \supop{\Omega}$, the operators defined in
\nec{2.58} take the form
\bes
T_{00} &= \supop{F_0} \mcirc \supop{U} \mcirc  \supop{H_0} =
\supop{U_{00}},
\quad
\\
\tT_{01} &= \supop{\tU_{01}},
\quad
T_{10} = \supop{U_{10}},
\quad
\tT_{11} = \supop{\tU_{11}},
\ees
where the notation $U_{00}$, $\tU_{01}$, $U_{10}$ and $\tU_{11}$ was
introduced in \nec{2.44}. For the reduction $\supop{\Omega} \cdot \supop{U}$
to exist, the series for $\sigma_1$ in \nec{2.22} must be absolutely
convergent for any $\rho_0 \in \cL(\cH_0)$. By linearity, it is sufficient to
show this for a $\rho_0$ of the form (a pure state)
\be
\rho_0 = \psi_0 \otimes \psi_0^\dagger
:= \supop{\psi_0} ,
\qquad
 \psi_0 \in \cH_0
\ee
where $\psi_0^\dagger$ is the canonical form dual to $\psi_0$ (under the
Hilbert space scalar product), and we have introduced the notation
$\supop{\psi} := \psi \otimes \psi^\dagger$ for vectors, then
\be
\| \supop{\psi} \|_t  = \| \psi \|^2
.
\ee
For those $\rho_0$ the relation
\be
\sigma_1^{(n)} =
\tT_{11}^{n-1} T_{10} (\rho_0)
\quad \forall n \ge 1
\ee
becomes
\be
\sigma_1^{(n)} =
\supop{\tU_{11}^{n-1} U_{10}  \psi_0 }
=
\supop{ \phi_1^{(n)} }
\quad \forall n \ge 1
,
\ee
where $\phi_1^{(n)}$ is that of \nec{1.53}. Hence 
\be
\| \sigma_1^{(n)} \|_t = \| \phi_1^{(n)} \|^2 .
\ee
The absolute convergence of the quantum channels series then follows from that
already established for the reduction of unitary operators. This shows that
$\supop{\Omega}\cdot\supop{U}$ exists.

Along the same lines, it is straightforward to verify that
\be
\sigma_0^{(n)} = \supop{ \phi_0^{(n)} } = \supop{A_n} (\rho_0)
\quad \forall n \ge 1
.
\ee
This implies that
\be
\supop{\Omega}\cdot\supop{U}
= \sum_{n=1}^\infty \supop{A_n} =
T_{U,\Omega}
.
\ee
\qed

The following nice relations have then been obtained:
\be
\Omega \cdot U = \sum_{n=1}^\infty A_n
,
\qquad
\supop{\Omega} \cdot \supop{U} = \sum_{n=1}^\infty \supop{A_n}
\,.
\ee

Proposition \ref{pr:12} will be invoked in Sec. \ref{sec:4}.

Let $\hat{\cV} \subseteq \cL(\cF_1)$ denote the largest positive subspace
which is invariant under $T' \mcirc R$.  By {\em positive subspace} we mean
spanned by positive operators. The invariance implies
$\tT_{11} \hat{\cV} \subseteq\hat{\cV} $ and $\tT_{01} \hat{\cV} = 0$, and
also that $\tT_{11}$ is a quantum channel when restricted to $\cV$.  Clearly,
if some non-null component of $\sigma_1^{(n)}$ ever enters into $\hat{\cV}$,
such component will remain there and the series will not converge. To be more
specific, let us assume that for some $n$ during the iteration \nec{2.22},
\be
\sigma_1^{(n)} = \sigma_W + \sigma_V,
\qquad
\sigma_W \ge 0 ,
\quad
\sigma_V \ge 0 ,
\qquad
0 \neq \sigma_V \in \hat{\cV}
.
\ee
Then $\| \sigma_V \|_t > 0$. Since $\tT_{11}$ is CP in $\cL(\cF_1)$ and
CPTP within $\hat{\cV}$,
\bes
\| \sigma_1^{(n+k)} \|_t
&=
\| \tT_{11}^k (\sigma_W) \|_t +
 \| \tT_{11}^k (\sigma_V) \|_t
 \\
 &=
\| \tT_{11}^k (\sigma_W) \|_t +
 \| \sigma_V \|_t
 \ge
 \| \sigma_V \|_t
 \quad
 \forall k\ge 0
.
\ees
It follows that the series of $\sigma_1$ will not converge.

\newlemma{lm:5} ~When $\cL(\cF_1)\neq 0$, $\hat{\cV} = 0$ if and only if
$\|(\tT_{11})^d \|_t < 1$, where $d:= \dim\cL(\cF_1)$.

\proof ~Suppose $\hat{\cV} \neq 0$. The superoperators $\tT_{11}^n$, $n\ge 0$,
are all quantum channels when restricted to $\hat{\cV}$, then
$\| (\tT_{11})^n \|_t = 1$ for all $n\ge0$. Now suppose
$\| (\tT_{11})^d \|_t = 1$. Then $\exists \sigma^{(0)} \in \cL(\cF_1)$,
$\sigma^{(0)}\ge 0$, $\Tr(\sigma^{(0)}) =1$, such that
$\Tr(\tT_{11}^d(\sigma^{(0)})) = 1$. Let
$\sigma^{(k)}:=\tT_{11}^k(\sigma^{(0)})$, $0\le k \le d$.  Since
$\|\tT_{11}\|_t \le 1$, necessarily $\Tr(\sigma^{(k)}) = 1$, and
$\sigma^{(k)} \ge 0$ for $0\le k \le d$. As no $d+1$ operators can be
linearly independent in $\cL(\cF_1)$, $\sigma^{(k_0)}$, for some
$0< k_0 \le d$, must be a linear combination of the previous $\sigma^{(k)}$
($0\le k< k_0$). Such $k_0$ positive operators span a non-zero invariant and
positive subspace of $\cL(\cF_1)$, hence $\hat{\cV} \neq 0$.  \qed

\newcorollary{cor:2} $\hat{\cV}=0$ is a sufficient condition for $R \cdot T$
to exist.

\proof ~If $\cL(\cF_1)=0$ the series \nec{2.22} converges in one step. If
$\cL(\cF_1) \neq 0$ and $\hat{\cV} = 0$, Lemma \ref{lm:5} implies that
$\| (\tT_{11})^d\|_t < 1$ for a $d \ge 1$, hence $R \cdot T$ exists from
Theorem \ref{th:7} with $\hat{\cW} = \cL(\cF_1)$. \qed

We have noted that the reduction of quantum channel with a subchannel does
not always exist (e.g. case $\bar{p}_{11}=1$ with $\bar{p}_{00}<1$ in Example \ref{ex:2}). However, the possibility that whenever \eqs{2.12c} have a
solution (perhaps with additional conditions) such solution must be a quantum
channel, has not been discarded. This issue is not settled in this work.

\section{ Implementation in quantum computation }
\label{sec:4}

\begin{figure}[ht]
  \begin{center}
    \includegraphics[height=60mm]{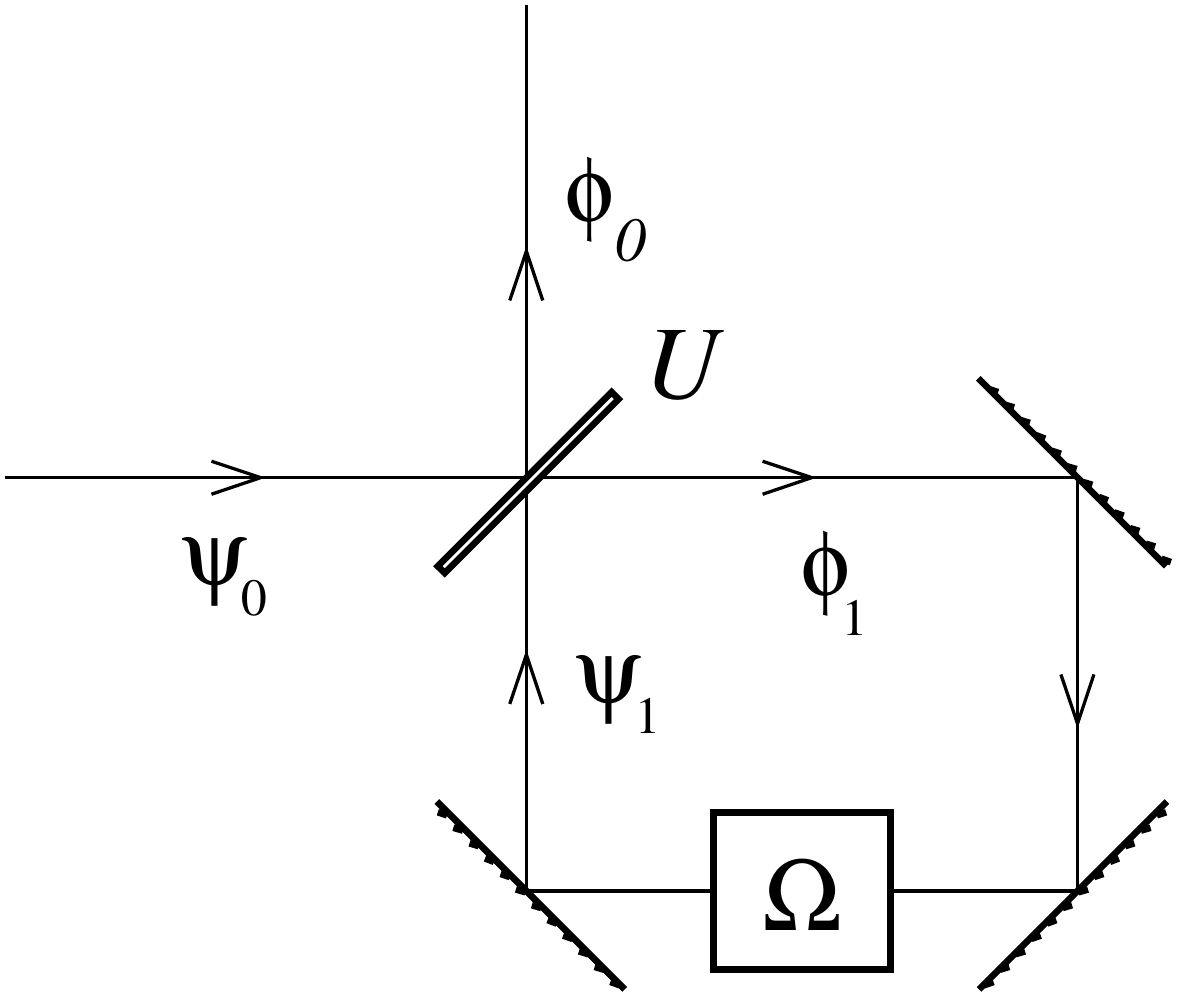}
    \end{center}
    \caption{ A unitary reduction implemented through a beam of stationary
      coherent light. }
\label{fig:4}
\end{figure}

Fig. \ref{fig:4} illustrates a unitary reduction implemented through a beam
of stationary coherent light. We disregard the polarization degree of freedom
as it does not play a role in this example. The incident beam with amplitude
$\psi_0$ enters an ideal beam splitter with transmission and reflection
coefficients $t$ and $r$, respectively, and associated unitary operator
$\ds U = \PM{ t & r' \\ r & t' }$. Part of the amplitude of
the beam is immediately reflected in the beam splitter (adding to $\phi_0$)
and part transmitted (adding to $\phi_1$). The latter component, after
reflecting on the mirrors returns to the splitter, again with the
possibilities of transmission (to $\phi_0$) or reflection (to $\phi_1$). A
phase shifter $\Omega$ can be inserted in the loop. The equations constraining
the amplitudes are then
\be
\psi_1 = \Omega \phi_1 ,
\qquad
\phi_1 = t \psi_0  + r' \psi_1
,
\qquad
\phi_0 = r \psi_0  + t' \psi_1
.
\ee
The solution produces a linear relationship $\phi_0 = S \psi_0$ where
$S= \Omega\cdot U$ is a phase identical to that already found in the Example
\ref{ex:1} with obvious identifications. The present illustration would
correspond to a space $\cH$ of a qubit (since the amplitude can arrive from
two different directions to the beam splitter) with a decomposition
$\C^2= \C \oplus \C$. Clearly an important problem arises if the setting is
directly applied to a photon, due to the delay between the various components
$\phi_0^{(n)}$ coming out of the splitter.

In all our previous discussions, with the exception of Proposition \ref{pr:5},
we have manipulated Hilbert spaces additively (direct sum of spaces) instead
of multiplicatively (tensor product of spaces). The latter is characteristic
of quantum computation, and is the reason of the efficiency of such approach
when treating certain computational problems. The additive structure is
natural for quantum graphs. There the dimension of the space increases
linearly with the number of external lines of the graph. In the case of
quantum circuits, the dimension increases exponentially with the number of
qubits.  It should be quite clear then that the reduction discussed in this
work does not refer to closing a loop of one or more qubit in a quantum
circuit. The reduction removes summands from a space rather than factors.

\begin{figure}[ht]
  \begin{center}
    \includegraphics[width=85mm]{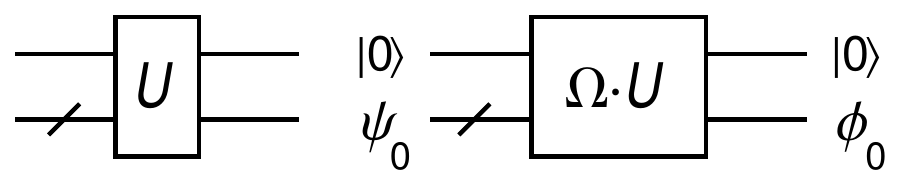}
    \end{center}
\caption{ A setting for unitary reduction in quantum circuits. }
\label{fig:5}
\end{figure}

Instead, a possible realization of the reduction with qubits is to consider
a system composed of a (control) qubit and another sector (composed by a
finite number of additional qubits) with Hilbert spaces $\C^2$ and $\cS$,
respectively. The total space is $\cH = \cF = \C^2 \otimes \cS$. If $\ket{0}$
and $\ket{1}$ form the computational basis of $\C^2$, the (additive)
decomposition $\cH = \cH_0 \oplus \cH_1$ comes out as
$\cH_0 =\cF_0 = \ket{0}\otimes \cS$ and $\cH_1=\cF_1 = \ket{1}\otimes
\cS$. The operator $U$ acts on the full space $\C^2 \otimes\cS$ (control qubit
plus system) $\Omega$ acts only on the sector $\ket{1}\otimes \cS$, while
$\Omega\cdot U$ acts on $\ket{0}\otimes \cS$. The setting is displayed in
Fig. \ref{fig:5}.

The question arises whether and how, for given $U$ and $\Omega$, the gate
$\Omega\cdot U$ can be constructed. Of course, as any other finite-dimensional
unitary operator, once the operator corresponding to $\Omega\cdot U$ has been
identified, it can be reproduced by a suitable circuit using a finite number
of one- and two-qubit gates. But the challenge is to do it using gates for $U$
and $\Omega$ as-given, i.e., as black boxes, without knowing how they act. The
answer is likely negative. Even the simpler gate $C(V)$ (controlled-$V$ gate,
the unitary operator $V$ acting on $\cS$) with black box $V$, cannot be
constructed with circuits \cite{Nielsen:2012yss}. The reason is of course that
if $V$ is inserted in a circuit producing a final unitary operator $W$, such
$W$ transforms covariantly under a phase change, that is, as $W\to \omega W$
when $V\to \omega V$ ($\omega\in \C, |\omega|=1$) while
$C(V) = P_0 \otimes I_\cS + P_1 \otimes V$ ($P_{0,1}$ denoting the projectors
on the $0$ and $1$ states of the control qubit) does not transform
covariantly.

In a $C(V)$ gate, $V$ acts potentially zero or one times. In attempting to
construct $\Omega \cdot U$ one finds an additional issue, namely, that the
gates $U$ and $\Omega$ (or actually their composition $U\Omega$) would need to
act repeatedly, in a loop, to reproduce the iterative procedure in \eq{1.53}
or equivalently the series in \nec{1.1c}. But even allowing loops and
controlled gates the construction is not obvious, nor perhaps viable.

\begin{figure}[ht]
  \begin{center}
    \includegraphics[height=25mm]{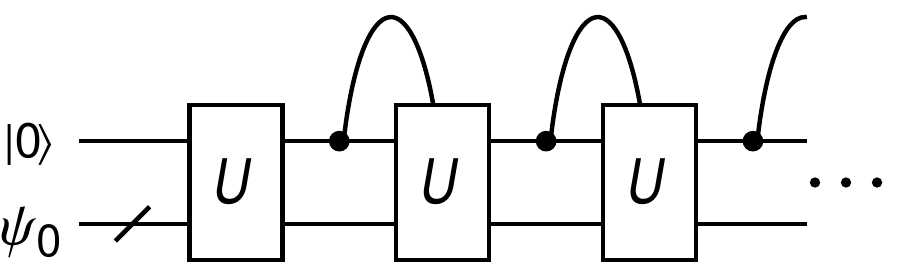}
    \end{center}
    \caption{ Implementation of \eq{1.38a} in quantum circuits with
      $\Omega=I$. It uses an unphysical controlled gate since the control
      qubit is also in the target.      
    }
\label{fig:6}
\end{figure}

A concrete construction of $\Omega\cdot U$ (assuming $\Omega=I$ for
simplicity) is illustrated in Fig. \ref{fig:6} which is a literal translation
of \eq{1.38a}: first $U$ acts on $\ket{0} \otimes \psi_0$ producing
$\ket{0} \otimes \phi_0^{(1)} + \ket{1} \otimes \phi_1^{(1)}$, then the
component with $\ket{0}$ is retained as such, while the component along
$\ket{1}$ passes through $U$ again and the algorithm is repeated. The
selection between the two components is effected by means of a unusual
controlled-$U$ gate, namely, $P_0 \otimes I_{\cS} + U P_1 \otimes I_{\cS}$.
It differs from a standard controlled gate in that here $U$ acts not only on
the target system $\cS$ but also on the control sector, so to say, the control
qubit controls itself. Leaving aside the infinite loop problem, the trouble is
that such controlled gate, although is a perfectly well-defined operator, is
not unitary, and not even invertible in general, and therefore it cannot be
implemented physically.

\begin{figure}[ht]
  \begin{center}
    \includegraphics[width=85mm]{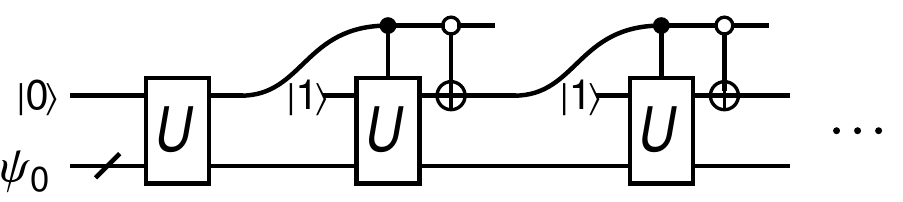}
    \end{center}
    \caption{ Exotic construction aiming at producing $\Omega\cdot U$ (for
      $\Omega=I$) using multiple control qubits. It actually produces a
      mixture of the $\phi_0^{(n)}$ instead of its coherent sum.}
\label{fig:7}
\end{figure}

A more exotic construction (again with $\Omega=I$) is attempted in
Fig. \ref{fig:7}. There, new ``control qubits'' in state $\ket{1}$ are being
added repeatedly. As it should, the component of a outgoing state with
$\ket{0}$ in the control qubit is never again modified while the component
with $\ket{1}$ is sent to $U$ to be processed over. Besides being awkward,
this construction (and similar ones) does not achieve its goal. The correct
states $\phi_0^{(n)}$ are produced, however they do not appear in the form
$\sum_n \phi_0^{(n)}$ but rather as $\sum_n u_n \otimes \phi_0^{(n)} $, where
the $u_n$ are orthonormal states of the control multiqubits. In practice this
is a mixture rather than a coherent sum. Note that since the states $u_n$ are
orthonormal it is not possible to unitarily rotate them to produce a product
state with a factor $\phi_0 = \sum_n \phi_0^{(n)}$.\footnote{Again, an
  unphysical non-injective map $u_n \to \ket{0}$ would produce the desired
  result.}

\begin{figure}[ht]
  \begin{center}
    \includegraphics[height=50mm]{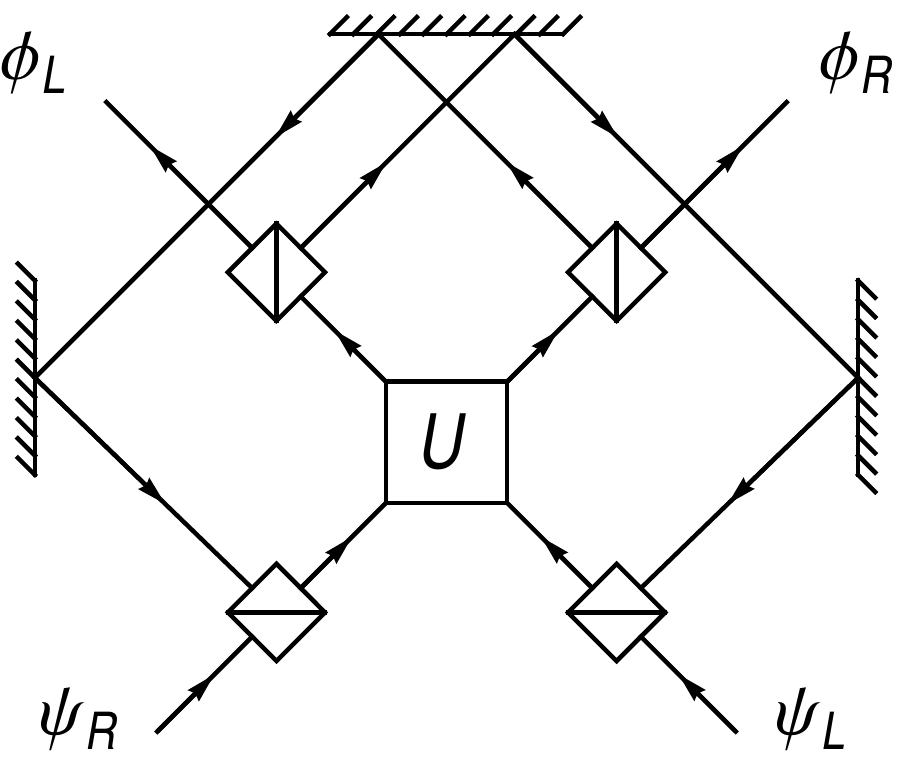}
    \end{center}
    \caption{ Scheme for an optical implementation of $\Omega \cdot U$ (with
      $\Omega=I$) for a black box gate $U$. The control qubit is the photon
      polarization $\ket{0}=\ket{H}$ or $\ket{1}=\ket{V}$, while $\cS =\C^2$
      is the photon momentum state $\ket{R}$ or $\ket{L}$. Both the input and
      the output are in the state $\ket{H}$.  }
\label{fig:8}
\end{figure}

While a gate $C(V)$ cannot be implemented for black box $V$ by means of
circuits, it can be implemented using other approaches, namely, optically
\cite{Araujo:2014rps}. The same idea can be attempted for $\Omega \cdot U$
with unknown $U$ and $\Omega$. Fig. \ref{fig:4} is already an instance and in
Fig. \ref{fig:8} a more general construction is displayed: The physical basis
is a photon which carries two qubits, one attached to polarization, $H$ or
$V$, and another to momentum state, $R$ or $L$. The polarization is the
control qubit and we identify $H$ with $0$ and $V$ with $1$, while $\cS =\C^2$
is spanned by $\ket{R}$ and $\ket{L}$. Hence the photon arrives in a state
$\psi_0 = \psi_R \ket{H,R} + \psi_L \ket{H,L} $ and leaves in a state
$\phi_0 = \phi_R \ket{H,R} + \phi_L \ket{H,L} $. It is assumed that the
polarizing beam splitters are such that perfectly transmit the horizontal
polarization and perfectly reflect the vertical one.

Regrettably, the construction in Fig. \ref{fig:8} works correctly for a beam
of light in a stationary state, however, its performance for actual photons is
deficient, due to the delay problem noted before. The states $\phi_0^{(n)}$
will be produced, but at different times $t_n = t_0 + n \tau$, $\tau$ being
the time needed to travel one loop (and sending new delayed photons all in the
same state would not solve the problem). Again, one obtains a state
$\sum_n u_n \otimes \phi_0^{(n)} $ where $u_n$ marks the number of loops
traveled before leaving the system, instead of a coherent sum of the states
$\phi_0^{(n)}$.

Another possibility, not aiming at constructing the reduction
$\Omega\cdot U$, is to measure the control qubit in the computational basis
(i.e., the observable $Z$) after each pass through the gate $U$, starting from
a normalized state $\ket{0}\otimes\psi_0$. If the result is $Z=-1$, $\Omega$
is applied and the state is sent again to $U$. The iteration ends the first
time that the result is $Z=+1$, and the state obtained is the output. This
produces a mixed state
$\sigma_0 = \sum_n \phi_0^{(n)} \otimes \phi_0^{(n)}{}^\dagger$. The map
$\rho_0 = \psi_0 \otimes \psi_0^\dagger \to \sigma_0$ is just the quantum
channel $\supop{\Omega}\cdot\supop{U}$ of Proposition \ref{pr:12}.

A similar mechanisms is available for the implementation of the reduction
$ R \cdot T $ of more general quantum channels. \Eq{2.18} can also be written
as
\be
R \cdot T =
\lim_{n\to\infty} \tT^n \mcirc (T_{00} + T_{10}),
\qquad
\tT = \supop{F_0} + \tT_{01} + \tT_{11}
\,.
\label{eq:2.18a}
\ee
All expressions involve directly $T$ and $R$ (a separate construction of $T'$
is not needed).  Nothing prevents a straightforward application of the
prescription indicated here: $T$ is applied to $\rho_0$, then the measurement
with projection-valued measure $F_0 + F_1 = I_\cF$ is carried out. If the
result is $F_0$ the process stops, otherwise $T \mcirc R$ is applied and the
measurement repeated. This is iterated until eventually the result is
$F_0$. This will happen eventually (that is, with probability $1$) if the
reduction exists. The density matrix $\sigma_0$ so obtained implements the
reduction.

\acknowledgments This work has been partially supported by
MCIN/AEI/10.13039/501100011033 under grant PID2020-114767 GB-I00, and by the
Junta de Andaluc{\'\i}a under grant No. FQM-225, and Spanish Ministerio de
Ciencia, Innovacion y Universidades under grant PID2023-147072NB-I00.



\begin{thebibliography}{99}

\bibitem{Kummerer:1987}
  B.~Kummerer and H.~Maassen,
  {\em The Essentially Commutative Dilations of Dynamical
    Semigroups on $M_n$,}\/
    Commun. Math. Phys. \textbf{109} (1987) 1
 
\bibitem{Audenaert:2008}
  K.~M.~R.~Audenaert and S.~Scheel,
  {\em On random unitary channels,}\/
  New J. Phys. \textbf{10} (2008) 023011

\bibitem{Kottos:1999}
T.~Kottos and U.~Smilansky,
{\em Periodic Orbit Theory and Spectral Statistics for Quantum Graphs,}\/
Annals Phys. \textbf{274} (1999), 76-124

\bibitem{Kuchment:2008dub}
P.~Kuchment,
{\em Quantum graphs: An Introduction and a brief survey,}\/
Proc. Symp. Pure Math. \textbf{77} (2008), 291-314

\bibitem{Berkolaiko:2013}
G.~Berkolaiko, and P.~Kuchment,
{\em Introduction to Quantum Graphs,}\/
Mathematical Surveys and Monographs, volume 186 (American Mathematical
Society, Providence, 2013).

\bibitem{Caudrelier:2009ay}
V.~Caudrelier and E.~Ragoucy,
{\em Direct computation of scattering matrices for general quantum graphs,}\/
Nucl. Phys. B \textbf{828} (2010), 515-535
  
\bibitem{Cai:2019}  
Z.~Cai and S.~C.~Benjamin,
{\em Constructing smaller Pauli twirling sets for arbitrary error channels,}\/
Scientific reports \textbf{9} (2019), 1–11 

\bibitem{Paulsen:2003}
  V.~Paulsen,
  {\em Completely bounded maps and operator algebras,}\/
  Cambridge  Studies in  Advanced  Mathematics  78  (Cambridge  University
  Press,  Cambridge, 2003).

\bibitem{Nielsen:2012yss}
M.~A.~Nielsen and I.~L.~Chuang,
{\em Quantum Computation and Quantum Information,}\/
Cambridge University Press, 2012,

\bibitem{Araujo:2014rps}
M.~Ara\'ujo, A.~Feix, F.~Costa and \v{C}.~Brukner,
{\em Quantum circuits cannot control unknown operations,}\/
New J. Phys. \textbf{16} (2014) no.9, 093026

\end{thebibliography}
\end{document}